\documentclass[twocolumn]{aastex62}

\usepackage{xspace}
\usepackage{graphicx}
\usepackage{times}
\usepackage{natbib}
\usepackage{amsmath,amsfonts,amssymb}
\usepackage{gensymb}
\usepackage{bm}
\usepackage{hyperref}
\usepackage{url}
\usepackage[caption=false]{subfig}
\usepackage{microtype}
\usepackage{rotating}
\usepackage{booktabs}
\usepackage{threeparttable}
\usepackage{tabularx}
\usepackage{listings}

\newcommand{\project}[1]{\textsl{#1}\xspace}

\newcommand{\rxte}{\project{RXTE}\xspace}
\newcommand{\nustar}{\project{NuSTAR}\xspace}
\newcommand{\Msun}{\ensuremath{M_\odot}\xspace}

\newcommand{\zsq}{\ensuremath{Z^2_n}\xspace}
\newcommand{\python}{\texttt{Python}\xspace}
\newcommand{\stingray}{\texttt{stingray}\xspace}
\newcommand{\astropy}{\texttt{astropy}\xspace}
\newcommand{\lightcurve}{\texttt{Lightcurve}\xspace}
\newcommand{\eventlist}{\texttt{EventList}\xspace}
\newcommand{\crossspectrum}{\texttt{Crossspectrum}\xspace}
\newcommand{\powerspectrum}{\texttt{Powerspectrum}\xspace}
\newcommand{\likelihood}{\texttt{Likelihood}\xspace}

\newcommand{\hendrics}{\texttt{HENDRICS}\xspace}
\newcommand{\dave}{\texttt{DAVE}\xspace}

\accepted{28 May 2018}
\submitjournal{ApJ}

\shorttitle{\stingray: A Modern \python\ Library For Spectral Timing}
\shortauthors{Huppenkothen et al.}

\begin{document}

\title{\stingray: A Modern \python\ Library For Spectral Timing}



\correspondingauthor{Daniela Huppenkothen}
\email{dhuppenk@uw.edu}

\author[0000-0002-1169-7486]{Daniela Huppenkothen}
\affil{DIRAC Institute, Department of Astronomy, University of Washington, 3910 15th Ave NE, Seattle, WA 98195}
\nocollaboration

\author[0000-0002-4576-9337]{Matteo Bachetti}
\affiliation{INAF-Osservatorio Astronomico di Cagliari, via della Scienza 5, I-09047 Selargius (CA), Italy}
\nocollaboration

\author[0000-0002-5041-3079]{Abigail L. Stevens}
\affiliation{Department of Physics \& Astronomy, Michigan State University, 567 Wilson Road, East Lansing, MI 48824, USA}
\affiliation{Department of Astronomy, University of Michigan, 1085 South University Avenue, Ann Arbor, MI 48109, USA}
\nocollaboration

\author{Simone Migliari}
\affiliation{ESAC/ESA, XMM-Newton Science Operations Centre, Camino Bajo del Castillo s/n, Urb. Villafranca del Castillo, 28692, Villanueva de la Cañada, Madrid, Spain.}
\affiliation{Timelab Technologies Ltd., 20-22 Wenlock Road, London N1 7GU, United Kingdom.}
\nocollaboration

\author{Paul Balm}
\affiliation{Timelab Technologies Ltd., 20-22 Wenlock Road, London N1 7GU, United Kingdom.}
\nocollaboration

\author{Omar Hammad}
\affiliation{AinShams University, Egypt}
\nocollaboration

\author{Usman Mahmood Khan}
\affiliation{Department of Computer Science, North Carolina State University, Raleigh, USA}
\nocollaboration

\author{Himanshu Mishra}
\affiliation{Indian Institute of Technology, Kharagpur West Bengal, India 721302}
\nocollaboration

\author{Haroon Rashid}
\affiliation{National University of Sciences and Technology (NUST), Islamabad 44000, Pakistan}
\nocollaboration

\author{Swapnil Sharma}
\affiliation{Indian Institute of Technology Mandi, Mandi, Himachal Pradesh, India}
\nocollaboration

\author[0000-0002-2985-6369]{Evandro Martinez Ribeiro}
\affiliation{Kapteyn Astronomical Institute, University of Groningen, P.O. Box 800, NL-9700 AV Groningen, The Netherlands.}
\nocollaboration

\author{Ricardo Valles Blanco}
\affiliation{Timelab Technologies Ltd., 20-22 Wenlock Road, London N1 7GU, United Kingdom.}
\nocollaboration



\begin{abstract}
This paper describes the design and implementation of \stingray, a library in \python built to perform time series analysis and related tasks on astronomical light curves. 
Its core functionality comprises a range of Fourier analysis techniques commonly used in spectral-timing analysis, as well as extensions for analyzing pulsar data, simulating data sets, and statistical modeling. 
Its modular build allows for easy extensions and incorporation of its methods into data analysis workflows and pipelines. We aim for the library to be a platform for the implementation of future spectral-timing techniques.  
Here, we describe the overall vision and framework, core functionality, extensions, and connections to high-level command-line and graphical interfaces.
The code is well-tested, with a test coverage of currently 95\%, and is accompanied by extensive API documentation and a set of step-by-step tutorials.
\end{abstract}



\section{Introduction} \label{sec:intro}

Variability is one of the key diagnostics in understanding the dynamics, emission processes and underlying physical mechanisms of astronomical objects. The flux of the majority of sources in the sky, from small asteroids to supermassive black holes, varies on time scales that can range from milliseconds to centuries, depending on the type of source.
The detection of periodic variations in the radio flux of certain celestial objects has led to the ground-breaking discovery of pulsars \citep{hewish1968}. Since this initial detection, the precise measurements of periods, spin-down effects and of intra-pulse variations in pulsars across the electro-magnetic spectrum from radio to gamma-rays has led to new insights on the structure of neutron stars and their magnetic fields \citep[e.g.][]{Lorimer2008,abdo2013,papitto2019}.
Similarly, accurate models of dips in stellar light curves have led to the discovery of thousands of exoplanets \citep[e.g.,][]{charbonneau2000,henry2000,coughlin2016}. In asteroseismology, the detection of oscillatory modes in the power spectra generated from the light curves of stars, including our sun, has allowed researchers a rare glimpse into the internal structure of these stars (see e.g. \citealt{dimauro2016} for a recent review), and Young Stellar Objects (YSOs) have been shown to be extremely variable (including rapid flaring) across the electromagnetic spectrum from radio to X-rays \citep[see e.g.][for a comprehensive analysis of YSOs in the Orion Nebula Cluster]{Forbrich_2017}, providing clues about the interaction of the forming stellar object and the circumstellar disk.
Methods very similar to those employed in asteroseismology are also used to study the interior of neutron stars and the dense matter equation of state through oscillations in magnetar bursts \citep[e.g.][]{huppenkothen2013} and giant flares \citep{israel2005,strohmayer2005,watts2006}. Analogous to studies of oscillations in magnetar giant flares and bursts, \citet{beloborodov2000} and \citet{guidorzi2016}, among others, have probed the variability of the prompt emission of Gamma-Ray Bursts in the quest to uncover the central engine of these sources, while in solar flares oscillations may give clues about the nature of the magnetic reconnection that powers these flares \citep{inglis2016}.

In a range of different astrophysical sources, including stars and compact objects, accretion plays a major role in their evolution and emission properties, giving rise to distinct patterns of brightness variations that allow us to study accretion physics in detail. For example, in white dwarfs, quasi-periodic variations attributed to magnetic gating of the accretion onto the star make it possible to measure the magnetic fields of these stars, even when they are too weak to be measured through other methods \citep{scaringi2017}. More generally, connections can be found between accretion onto young stellar objects and onto compact objects including supermassive black holes, linking physics across many different scales in mass and time \citep{scaringi2015}. Particularly in the study of black holes and neutron stars, the scientific developments of recent decades have brought a growing understanding that time and wavelength are intricately linked. 
Different spectral components react differently to changes in accretion rate and dynamics, leading to energy-dependent time lags, correlated variability, and higher-order effects \citep[for a review, see][]{uttley2014}. 

This has led to the study of accretion disks, in particular those of active galactic nuclei, via reverberation mapping \citep[e.g.,][]{blandford1982,Bentz2016}, and probes of the accretion disk geometry using the energy-dependence of quasi-periodic oscillations in stellar-mass black holes \citep[e.g.,][]{ingram2015,stevensuttley2016}. 
Understanding how the emission at various wavelengths changes with time is crucial for testing and expanding our understanding of general relativity in the strong-gravity limit, the dense matter equation of state and other fundamental questions in astrophysics.
In X-ray astronomy, there is now a wealth of public data sets of variable objects from missions such as the \textit{Rossi X-ray Timing Explorer}  (\textit{RXTE}; \citealt{Bradtetal93}), 
the X-ray Multi-Mirror Mission (\textit{XMM-Newton}; \citealt{jansen2001}), the \textit{Nuclear Spectroscopic Telescope Array} (\nustar; \citealt{nustar13}), \textit{Astrosat} \citep{singh2014},  and the \textit{Neutron Star Interior Composition Explorer} (\textit{NICER}; \citealt{gendreau2016}). 
In addition, planned missions such as the \textit{Advanced Telescope for High-Energy Astrophysics} \citep[\textit{Athena};][]{athenaXIFU} and proposed missions like the \textit{Enhanced X-ray Timing Polarimeter} \citep[\textit{eXTP};][]{extp16} and the \textit{Spectroscopic Time-Resolving Observatory for Broadband Energy X-rays} \citep[\textit{STROBE-X};][]{strobex18} will produce data sets of unprecedented size and complexity. 

Motivated by the ubiquity of (spectral) timing in astronomy, and the advent of these new data sets, we present \stingray, a well-tested, open-source \textit{Python} implementation of a range of core algorithms and methods used in time series analysis and spectral timing across the electromagnetic spectrum. The package has now been employed in a range of studies including radio observations of the galactic black hole X-ray binary Cygnus X-1 \citep{tetarenko2019}, optical and X-ray observations of accreting pulsars \citep{kennedy2018,jaisawal2018,brumback2018}, X-ray observations of accreting low-mass \citep{beri2019} and high-mass \citep{pike2019} X-ray binaries, Ultra-luminous X-ray Sources \citep{walton2018}, and statistical investigations of cospectra \citep{bachetti2017,huppenkothen2017}.

The paper layout is as follows: 
In Section \ref{sec:data}, we very briefly describe the data sets being used in this paper to showcase the implemented methods. 
In Section \ref{sec:vision}, we lay out the overall vision, followed by a description of the general package structure and the general development framework in Section \ref{sec:development}. 
The package's core functionality is shown in more detail in Section \ref{sec:core}, where we introduce basic classes for generating light curves and Fourier spectra of various types.
In Sections \ref{sec:modeling}, \ref{sec:simulator} and \ref{sec:pulsar}, we present the submodules enabling the statistical modeling of Fourier products, simulating light curves from stochastic processes, and pulsar analysis, respectively.
Sections \ref{sec:hendrics} and \ref{sec:dave} point to connections with a command-line interface and a graphical user interface which are being developed concurrently with \stingray. 
Finally, in Section \ref{sec:future} we lay out our future development plans. 
Note that we intentionally omit code examples and specific implementation details in this manuscript in order to preserve longer-term accuracy. 
All code to reproduce the figures in this paper is available online,\footnote{\url{https://github.com/StingraySoftware/stingraypaper}} as is a full suite of up-to-date tutorials.\footnote{\label{foot:nb}\url{https://github.com/StingraySoftware/notebooks}}

\section{Data}
\label{sec:data}
Throughout the paper, we use real observations of compact objects to demonstrate the functionality of the software in this package. 
In the following sections, we give brief introductions into the observations used and the data reduction processes applied before using the resulting event files and light curves with \stingray.

\subsection{GX~339--4}
\label{sec:gx339}
GX 339--4 is a stellar-mass black hole in a low-mass X-ray binary \citep{Hynesetal03}. 
The black hole has a lower mass limit of $\sim$\,7\Msun\ \citep{MunozDariasetal08} and possibly a near-maximal spin \citep{Ludlametal15}. 
The system also likely has a low binary orbit inclination; it has been constrained to $37 \degree <i\lesssim 60\degree$ from optical and X-ray observations \citep{Heidaetal17, Zdziarskietal98}, and spectral modeling by \citet{WangJietal18} estimates $i\approx 40\degree$.
We use an observation from the  \textit{RXTE} Proportional Counter Array (PCA; \citealt{Jahodaetal96}) in NASA's High Energy Astrophysics Science Archive Research Center (HEASARC) from the 2010 outburst of GX 339--4 \citep{Yamaokaetal10}, with the observation taken from UT 2010-04-22 23:36:52 to UT 2010-04-23 00:01:10 (observation ID 95409-01-15-06).
This observation was taken in 64-channel event mode with $122\,\mu$s time resolution (\texttt{E\_125us\_64M\_0\_1s}).
The following filtering criteria were used to obtain Good Time Intervals (GTIs): Proportional Counter Unit (PCU) 2 is on, two or more PCUs are on, elevation angle $>\,10\degree$, and target offset $<\,0.02\degree$. 
Time since the South Atlantic Anomaly passage was not filtered on. 
Applying these filters, we have $\sim$\,1\,ks of good data. 
Since the observation is short, the data were not barycentered\footnote{\label{foot:bary} ``Barycentering'' the data applies a spacecraft clock correction to correct the photon arrival times to the solar system barycenter. This is commonly done with the FTOOL \texttt{barycorr} from HEASoft.} before analysis.

\subsection{KIC12158940}
\label{sec:kepleragn}

KIC12158940 is one of 21 Active Galactic Nuclei studied in \citet{smith2018} and was observed for 12 epochs with the \textit{Kepler} Space Telescope \citep{borucki2010}. \citet{smith2018} report an average Kepler magnitude of 14.85 \citep[originally calculated in][]{brown2011} and a black hole mass of $\log{M_{\sun} }= 8.04$. The power spectrum shows a break at a characteristic time scale of $\tau_\mathrm{char} = 31.6$ and a steep power spectral slope of $\Gamma = -3.3$.

We downloaded the pre-produced light curves for all twelve epochs from the Barbara A. Mikulski Archive for Space Telescopes (MAST). We note that \citet{smith2018} caution that because Kepler was designed for observing stars, these pre-produced AGN light curves contain instrumental artifacts (e.g. too small extraction apertures and rolling band noise) that render these light curves too imprecise to derive scientific conclusions from them. Because the purpose of this work is to showcase \stingray's capabilities on different types of data rather than deriving physical properties, we continue with the pre-produced MAST light curves.

We did spot checks using a larger aperture to extract photometric fluxes from the full-frame images using the Python package \textit{Lightkurve} \citep{lightkurve} and find no discernible differences due to the smaller apertures used by the \textit{Kepler} team. 
For astrophysical investigations, however, we urge the reader interested analysing AGN light curves observed with Kepler to follow the methods laid out in \citet{smith2018} to produce de-biased light curves before using \stingray. We split the \textit{Kepler} light curves for KIC12158940 along data gaps to produce contiguous segments, and rebinned all segments to a resolution of 0.075 days in order to generate evenly sampled light curves.

\subsection{Hercules X-1}
\label{sec:herx1}

Hercules X-1 (Her X-1) is a well-known persistent X- ray binary pulsar with a period of $P = 1.23\,\mathrm{s}$ \citep{tananbaum1972} in a binary system with a $\sim$\,2.2\,$\Msun$ stellar companion HZ Herculis \citep{davidsen1972,forman1972,bahcall1972,reynolds1997,leahy2014} with an orbital period of $P_\mathrm{orb}=1.7\,\mathrm{days}$ and super-orbital variations on a $\sim$\,35-day timescale \citep{giacconi1973,scott1999,igna2011}. 
The companion's type varies between late-type A and early-type B with orbital phase \citep{anderson1994,cheng1995}. 
For this work, we considered two of the several observations of Her X-1 with \nustar. 
The first observation was taken from UT 2012-09-19 to UT 2012-09-20 and was one among several used by \citet{Fuerst13} to characterize the cyclotron resonance scattering features (CRSFs) in the spectrum of the source. 
The second was taken from UT 2016-08-20 to UT 2016-08-21 and was used by \citet{staubert_2017} to detect an inversion of the decay of the CRSF. 
We used observation IDs 30002006002 and 10202002002 from HEASARC and barycentered$^{\ref{foot:bary}}$ the data with the latest (as of Nov. 27, 2018) \nustar clock correction file.
For our analysis, we considered photons from 3 to 79 keV at most 50$^{\prime\prime}$ from the nominal position of the source, extracted from the two identical Focal Plane Modules A and B (FPMA and FPMB, respectively) onboard the spacecraft. 
We used a total of 32.67\,ks of good data in the first observation and 36.56\,ks in the second, only selecting intervals longer than 10s.


\section{Vision and General Package Framework}
\label{sec:vision}

Despite decades of research, the field of spectral timing in high-energy astrophysics is fragmented in terms of software; there is no commonly accepted, up-to-date framework for the core data analysis tasks involved in (spectral) timing. 
Code is often siloed within groups, making it difficult to reproduce scientific results. 
Additionally, the scarcity of fully open-source tools constitutes a significant barrier to entry for researchers new to the field, since it effectively requires anyone not part of collaborations with an existing private code base to write their own software from scratch. 
The NASA library \texttt{xronos} is, to our knowledge, the only widely used open-source library in this field, and has several shortcomings. 
In particular, it performs only a few of the most basic tasks, and it has not been maintained since 2004. 
Other open-source projects use languages that either require an expensive license (e.g., IDL) or have a limited scope (e.g., S-Lang).
This dearth of  software for spectral timing motivated the development of \stingray,\footnote{\stingray was named partly in homage to the popular 1960s childrens' TV series, from which \stingray's motto derives: \textit{Anything can happen in the next half hour (including spectral timing made easy)!}} a library built entirely in the widely-used \python language and based on \astropy functionality. 
\stingray aims to make many of the core Fourier analysis tools used in timing and spectral-timing analysis available to a large range of researchers while providing a common platform for new methods and tools as they enter the field. 

It includes the most relevant functionality in its core package, while extending that functionality in its subpackages in several ways, allowing for easy modeling of light curves and power spectra, simulation of synthetic data sets and pulsar timing. 

Its core idea is to provide time series analysis methods in an accessible, unit-tested way, built as a series of object-oriented modules. 
In practice, data analysis requirements are varied and depend on the type of data, the wavelength the observation was taken at, and the object being observed. 
With this in mind, \stingray does not aim to provide full-stack data analysis workflows; rather, it provides the core building blocks for users to build such workflows themselves, based on the specific data analysis requirements of their source and observation. 
The modularity of its classes allows for easy incorporation of existing \stingray functionality into larger data analysis workflows and pipelines, while being easily extensible for cases that the library currently does not cover. 

\stingray separates out core functionality from several more specialized tasks based on those core classes and functions. Constructs related to data products as well as Fourier transforms of the data (e.g., power spectra, cross spectra, time lags, and other spectral timing products) are considered core functionality, as are some utility functions and classes, for example related to GTI calculations. 

This core functionality is extended in various ways in currently three subpackages. 
The \texttt{modeling} subpackage (see also Section \ref{sec:modeling}) provides a framework for modeling light curves and Fourier spectra with parametric functions. 
Based on this framework, it allows users to search for (quasi-)periodic oscillations in light curves with stochastic variability, and provides convenience functions to aid standard tasks like fitting Lorentzian functions to power spectra. 

The subpackage \texttt{simulator} (Section \ref{sec:simulator}) provides important functionality to allow efficient simulation of time series from a range of stochastic processes. This includes simulation of light curves from power spectral models as well as the use of transfer functions to introduce time lags and higher-order effects. 

Finally, the subpackage \texttt{pulsar} implements a range of methods particularly useful for period searches in pulsars.

\stingray is designed to be used both as a standalone package, and is also at the core of two other software packages currently under development: 
\hendrics and \dave. 
\hendrics (\citealt{hendrics}; see also Section \ref{sec:hendrics}) provides pre-built data analysis workflows using \stingray core functionality. 
These workflows are accessible from the command line and are provided for some common data types and data analysis tasks. 
\dave (see also Section \ref{sec:dave}) provides a Graphical User Interface on top of \stingray to allow for easy interactive exploratory data analysis.
 
As of v0.1, the core functionality of \stingray depends exclusively on \texttt{numpy} \citep{numpy}, \texttt{scipy} \citep{scipy} and \texttt{astropy} \citep{astropy}, with optional plotting functionality supplied by \texttt{matplotlib} \citep{matplotlib} . 
The \texttt{modeling} subpackage optionally uses sampling methods supplied by \texttt{emcee} \citep{emcee}, some functionality implemented in \texttt{statsmodels}, and plotting using \texttt{corner} \citep{corner}. 
The \texttt{pulse} subpackage optionally allows for just-in-time compilation using \texttt{numba} \citep{numba} for computational efficiency, and for advanced pulsar timing models using \texttt{PINT}\footnote{\url{https://github.com/nanograv/PINT}} .

This paper describes \stingray v0.1, released on 2018-02-12. 
As with most open-source packages, \stingray is under continuous development and welcomes contributions from the community, including suggestions for new subpackages to be implemented.


\section{Development and Integration Environment}
\label{sec:development}

\stingray is developed entirely in Python 3, with backwards compatibility to Python 2.7 where possible through the integration package \texttt{six}. 
Development is version-controlled through \texttt{git}, and officially hosted on GitHub through the organization \textit{StingraySoftware},\footnote{\url{https://github.com/StingraySoftware/}} where several interconnected repositories related to \stingray live, including the core library, extension packages \hendrics and \texttt{DAVE}, the suite of tutorials, the website, and this manuscript. 
All patches and code are submitted via pull requests to the \stingray repository, and checked by a maintainer for correctness of algorithms, adherence to standards of code, documentation, and tests. 
As an \textit{Astropy} Affiliated Package,\footnote{\url{http://www.astropy.org/affiliated/}} we follow the coding standards as well as community guidelines (including the Code of Conduct) set out by the \textit{Astropy} community. 
All code within the \stingray core library is subject to extensive unit testing, with compatibility across platforms as well as different versions of Python and required packages controlled through Continuous Integration services \textit{Travis} (Unix platforms) and \textit{AppVeyor} (Windows). 
Test coverage is checked using \textit{Coveralls}. 
All user-facing functions and classes within \stingray must have documentation in the form of docstrings, compiled and built along with the main documentation pages using \textit{sphinx} and hosted on \texttt{readthedocs}.\footnote{\url{https://stingray.readthedocs.io/en/latest/}}
Tutorials are provided in the form of executable \textit{Jupyter} notebooks in a separate repository,$^{\ref{foot:nb}}$ which can either be run interactively using \textit{Binder} \citep{project_jupyter-proc-scipy-2018} or viewed as part of the documentation.

\begin{figure*}[htbp]
\begin{center}
\includegraphics[width=\textwidth]{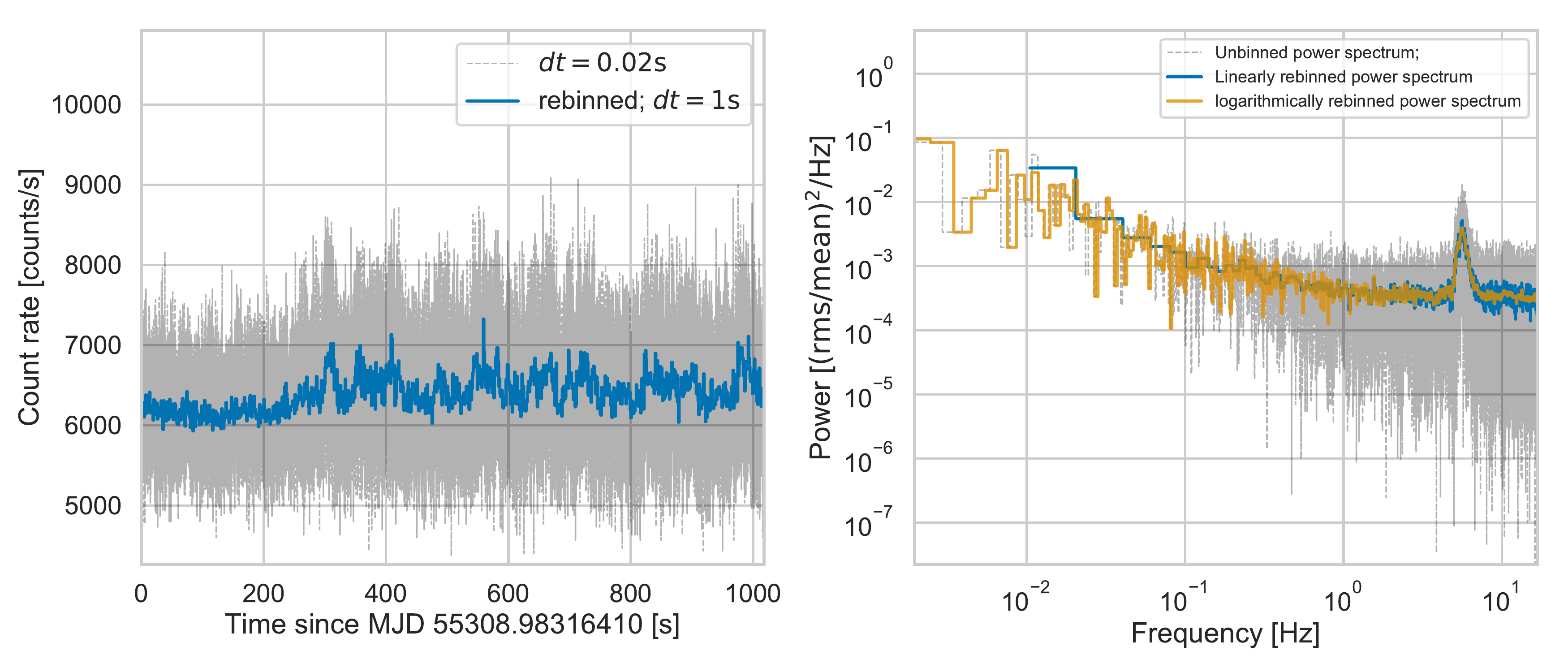}
\caption{Left panel: A $\sim 1 \mathrm{ks}$ \rxte\ observation of the black hole X-ray binary GX~339--4. 
Details of the observation can be found in Section \ref{sec:gx339}. 
In gray, we show the light curve produced by binning the events into $0.02\,\mathrm{s}$ bins. 
The blue line corresponds to the rebinned light curve at $dt = 1.0\,\mathrm{s}$. 
Right panel: we show the power spectrum calculated from the light curve in the left panel (gray), as well as a version of the same power spectrum that has been linearly rebinned (blue) and logarithmically rebinned (orange) in frequency.}
\label{fig:psd}
\end{center}
\end{figure*}

\begin{figure*}[htbp]
\begin{center}
\includegraphics[width=\textwidth]{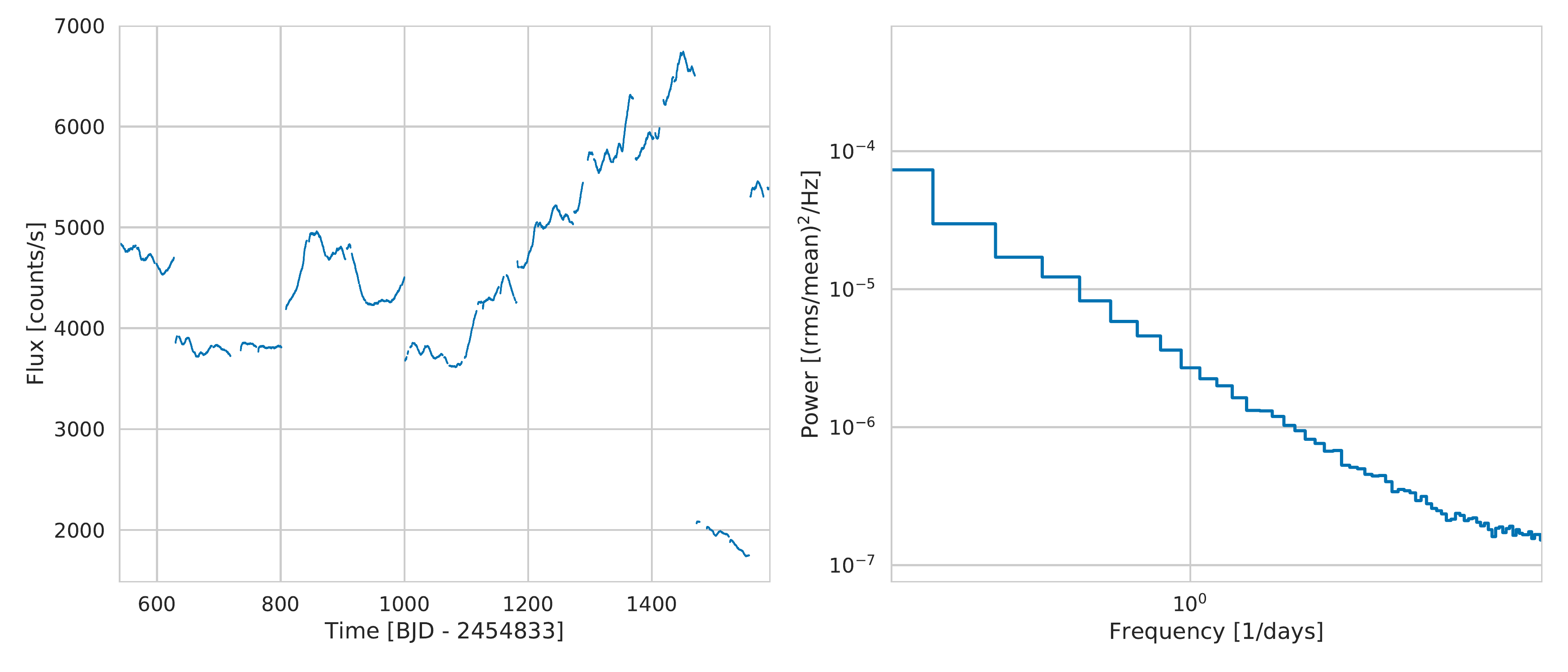}
\caption{Left panel: Light curve of all twelve epochs of the \textit{Kepler} observation of the AGN KIC12158940, rebinned to 0.075 days. 
Details of the observation can be found in Section \ref{sec:kepleragn}. While there are error bars on the \textit{Kepler} data points, they are so small on this scale as to be virtually invisible.
The right panel shows the corresponding averaged power spectrum, using a total of $50$ segments, each with a length of 10 days.}
\label{fig:kepleragn}
\end{center}
\end{figure*}

\section{Core Functionality}
\label{sec:core}

\stingray imports its core functions and classes from the top level package. 
These classes define the basic data structures such as light curves and cross- as well as power spectra that are used in much of the higher-level functionality provided in the sub-packages. 
Additionally, it incorporates a number of utilities for dealing with GTIs as well as input and output of data sets. 

\subsection{The \texttt{Lightcurve} class}
\label{sec:lightcurve}

We expect \stingray to be used largely on data sets of two forms: (1) event data (i.e., recordings of arrival times of individual photons) or (2) binned light curves (i.e. measurements of brightness in units of flux, magnitude or counts as a function of time). 

The majority of methods in \stingray use binned light curves, which we thus currently consider the default format. 
The \lightcurve class defines a basic data structure to store binned light curves. Its attributes include arrays describing time bins and associated (flux or counts) measurements, the number of data points in the light curve, the time resolution and the total duration of the light curve. 
For unevenly sampled light curves, the time resolution \texttt{dt} will be defined as the median difference between time bin midpoints. Users can pass uncertainties for measurements directly, or pass a string defining the statistical distribution of the data points for automatic calculation. 
By default, a Poisson distribution is assumed, appropriate for binned event data. 

There are two ways to generate a \texttt{Lightcurve} object: in the standard case, the instrument has recorded a binned time series of $N$ pairs of time stamps and count (rate) or flux values, $\{t_k, c_k \}_{k=1}^{N}$. 
In this case, one can simply instantiate a \texttt{Lightcurve} object with the keywords \texttt{time} and \texttt{counts} (and optionally set \texttt{use\_counts=False} when the input is in units of counts per second). 
In cases where the native data format is events (e.g., photon arrival times) it is possible to use the static method \texttt{Lightcurve.make\_lightcurve}, passing the array of events as well as a time resolution \texttt{dt} to create a new light curve from the events.

Various operations are implemented for class \lightcurve. 
Custom behaviour of the $+$ and $-$ operators allows straightforward addition and subtraction of light curves from one another. 
Assuming the light curves have the same time bins, the $+$ and $-$ operators will add or subtract the flux or counts measurements, respectively, and return a new \lightcurve\ object with the results. 
Other common operations implemented include time-shifting the light curve by a constant factor, joining two light curves into a single object, truncating a light curve at a certain time bin, and input/output operations to read or write objects from/to disk in various formats (HDF5, FITS and ASCII are currently supported). 
For light curves that do not have consecutive time bins, there is a sorting operation, as well as the option to sort the light curve by the ascending or descending flux or counts. 

We provide support for GTIs in many methods and implement rebinning the light curve to a new time resolution larger than the native resolution of the data (interpolation to a finer resolution is currently not supported). 
In Figure \ref{fig:psd} (left panel), we show an example observation of GX~339--4 as taken with with \rxte, and in Figure \ref{fig:kepleragn} (left panel) a \textit{Kepler} observation of KIC12158940.
\stingray implements basic methods for plotting (useful for a quick look at the data).

\subsection{The \texttt{Events} class}

At short wavelengths, data is largely recorded as \textit{photon events}, where arrival times at the detector are recorded for each photon independently, along with a number of other properties of the event (for example an energy channel in which the photon was recorded in, which can be transformed to a rough estimate of the energy of the original photon arriving at the detector).

Even for a single instrument, there are often multiple types of data that can be recorded, resulting in a plethora of data formats and internal schemas for how data is stored within the binary files distributed to the community. 
\stingray implements a basic \eventlist class that acts as a container for event data, but does not aim to encompass all data types of all current (and future) instruments. 
Instead, it aims to abstract away from instrument-specific idosyncrasies as much as possible and remain mission-agnostic. 
In its basic form, it takes arrays with time stamps and optionally corresponding photon energies as input, and implements a set of basic methods. 
Similarly to \lightcurve, it provides basic input/output (I/O) functionality in the form of \texttt{read} and \texttt{write} methods as well as a method to join event lists, which can be particularly useful when data is recorded in several independent detector, as is common for several current and future X-ray missions. 
The \verb|to_lc| method provides straightforward connection to create a \lightcurve directly out of an \eventlist object. 
In return, it is possible to create an \eventlist out of a \lightcurve object using the \verb|from_lc|. 
The latter will create $N_i$ events, each with a time stamp equation to the time bin $t_i$, where $N_i$ is the number of counts in bin $i$ (event lists are, by their very definition only a useful data product if the light curve used to simulate comes from photon counting data in the first place). 
It is possible to simulate more physically meaningful photon events from a given light curve and energy spectrum using the \verb|simulate_times| and \verb|simulate_energies| methods (from the \texttt{simulator} package, Section~\ref{sec:simulator}), which employ a combination of interpolation and rejection sampling to accurately draw events from the given light curve and spectrum.

\subsection{Cross Spectra and Power Spectra}
\label{sec:csps}

The cross spectrum and the power spectrum\footnote{In the signal processing literature, generally a distinction is made between the power spectrum, which
describes the process at the source generating variable time series, and the periodogram, which denotes a realization of said power spectrum, i.e., the time series we actually observe, which is an \textit{estimator} of the underlying process. While the products generated by \stingray are generally derived from data, and therefore periodograms, the astronomy literature usually denotes them by the term power spectrum. We follow this convention here as we do within the software package itself.} 
are closely related (for a pedagogical introduction into Fourier analysis, see \citealt{vanderklis1989}; see also \citealt{uttley2014} for a recent review of spectral timing techniques). 
Computing the cross spectrum requires two evenly sampled time series $\mathbf{y}_1 = \{y_{1,i}\}_{i=1}^{N} $ and $\mathbf{y}_2 =  \{y_{2,i}\}_{i=1}^{N}$ taken simultaneously at exactly the same time intervals $\{t_i \}_{i=1}^N$. 
Under this assumption, one may then compute the discrete Fourier transform of each time series, $\mathcal{F}_1$ and $\mathcal{F}_2$ independently, and multiply $\mathcal{F}_1$ with $\mathcal{F}^{*}_2$, i.e. the Fourier transform of $\mathbf{y}_1$ with the \textit{complex conjugate} of the Fourier transform of $\mathbf{y}_2$. 

Because the power spectrum is defined as the square of the real part of the Fourier amplitudes of a single, evenly sampled time series, it can be formulated as the special case of the cross spectrum where $\mathbf{y}_1 = \mathbf{y}_2$. 
In \stingray, we implement a class \crossspectrum, which takes two \lightcurve objects as input and internally calculates the complex cross spectrum in one of a number of common normalizations (see below).  Because many of the internal calculations are the same, the class \powerspectrum is implemented as a subclass of \crossspectrum, but takes only a single \lightcurve object instead of two. 

There are several popular normalizations for the real part of the cross spectrum as well as the power spectrum implemented in \stingray: the \textit{Leahy normalization} \citep{leahy1983} is defined such that for simple white noise, the power spectrum will follow a $\chi^2$ distribution with $2$ degrees of freedom around a mean value of $2$, and the cospectrum---the real part of the cross spectrum---will follow a Laplace distribution centred on $0$ with a scale parameter of $1$ \citep{huppenkothen2017}. It is particularly useful for period searches, because the white noise level is well understood and always the same (but be aware that detector effects like dead time can distort the power spectrum in practice; \citealt{Bachetti+15}).
For light curves with complex variability patterns, and especially for understanding how these patterns contribute to the overall variance observed, the \textit{fractional rms-squared normalization} \citep{belloni1990,miyamoto1992}  or the \textit{absolute rms-squared normalization} \citep{uttley2001} may be more appropriate choices. 

The classes \crossspectrum and \powerspectrum share most of the implemented methods, except where otherwise noted. 
Both classes include methods to rebin cross- and power spectra. Linear rebinning is implemented analogously to the method in class \lightcurve. 
Additionally, logarithmic binning is implemented in the method \texttt{rebin\_log} in such a way that the bin width at a given frequency increases by a fraction of the previous bin width:

\[
d\nu_{i+1} = d\nu_{i} (1 + f) \; ,
\]

\noindent where $f$ is some constant factor by which the frequency resolution increases, often $f = 0.01$. 

Classical period searches are often formulated as outlier detection problems from an expected statistical distribution. 
Assuming the signal is sufficiently coherent such that all of the signal power is concentrated in one bin, one may calculate the chance probability that an observed power in the spectrum was generated by statistical fluctuations alone. 
For the white noise case, the equations to accurately calculate a $p$-value of rejecting the hypothesis that a given outlier in the power spectrum was generated by noise are defined in \citet{Groth1975}, and can be calculated for one or multiple powers in a \powerspectrum object using the \verb|classical_significances| method, which enables computation of a (trial-corrected) $p$-value for a given power in the presence of white noise.
Note that the cross spectrum does not follow the same distribution \citep{huppenkothen2017}, and the recently derived statistical distributions for this case will be implemented in a future version of \stingray. 

In many practical applications, users may wish to average power- or cross spectra from multiple light curve segments in order to suppress statistical noise. 
This can be done with the appropriate classes \texttt{AveragedPowerspectrum} and \texttt{AveragedCrossspectrum}, which take a \lightcurve object or list of \lightcurve objects as an input and will compute averaged Fourier products by dividing the light curve into $N$ segments of a given size $\tau_\mathrm{seg}$. 
The Fourier spectra (either cross spectra or power spectra) are averaged together. 
Both are subclasses of \crossspectrum, and either inherit or override many of the methods relevant for those classes as well. 
Examples of the kinds of products produced by the classes and methods introduced above are given in Figures \ref{fig:psd} and \ref{fig:kepleragn} (right panels).

For averaged cross spectra, it is possible to calculate the time lag between variability in two simultaneous light curves, for example, if the two light curves cover different energy bands \citep{Vaughanetal94}. 
The time lag $\tau_j$ is defined as

\[
\tau_j = \frac{\phi_j}{2\pi\nu_j} \; 
\]

\noindent for a phase angle $\phi_j$ derived from the imaginary component of the complex cross spectrum, and a mid-bin frequency $\nu_j$. 
Similarly, it is possible to calculate the coherence from the cross spectrum \citep{vaughan1997,nowak1999}, defined as 

\begin{equation}
c_j = \frac{C_{xy,j}}{C_{x,j} C_{y,j}} \; . 
\end{equation}

\noindent Here, $C_{xy,j}$ corresponds to the real part of the unnormalized cross spectrum, and $C_{x,j}$ and $C_{y,j}$ correspond to the analogous squared amplitudes of the power spectrum for each individual light curve. 
The error on $\tau_j$ and $c_j$ are also computed in \stingray.

\begin{figure}[htbp]
\begin{center}
\includegraphics[width=9cm]{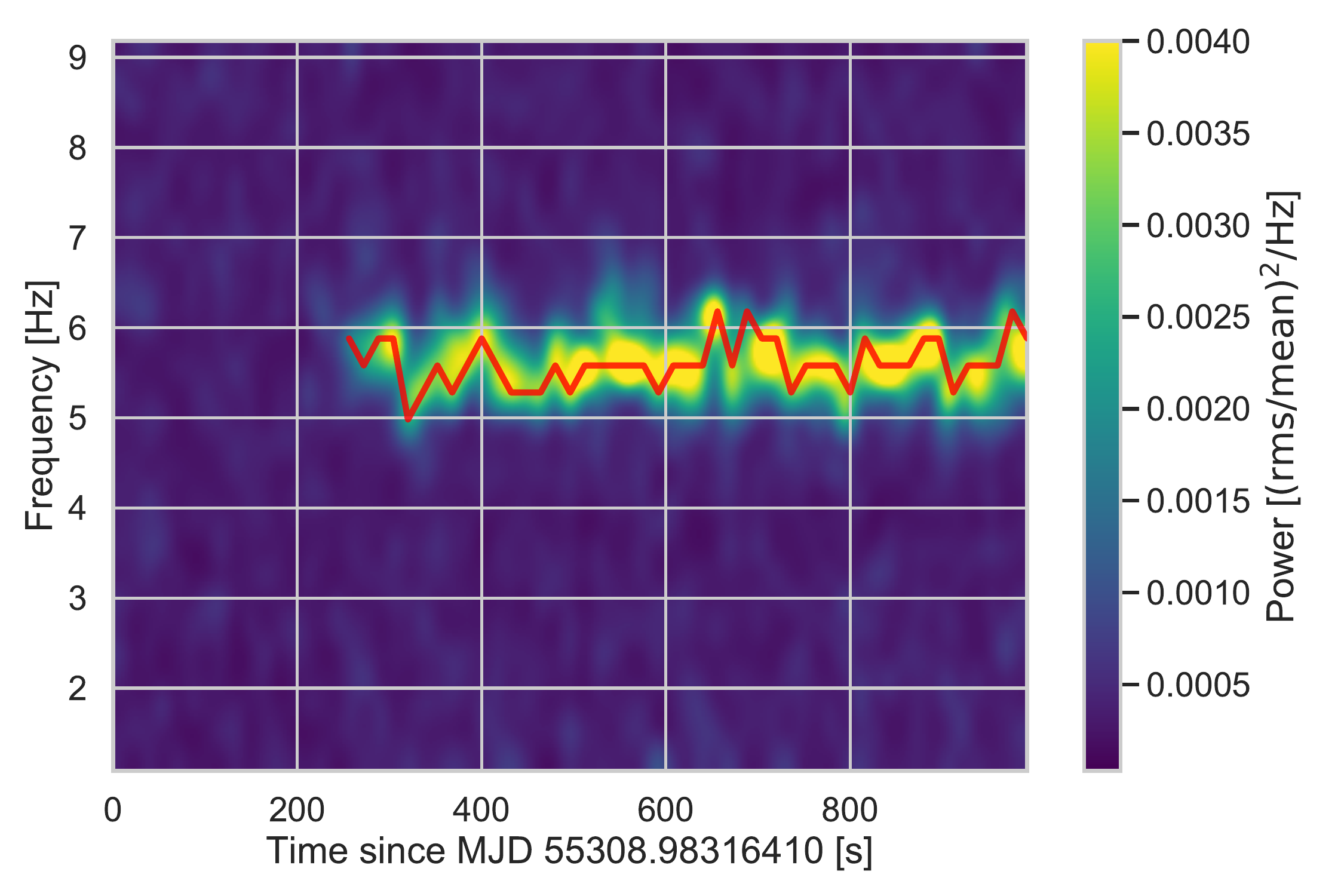}
\caption{An example of a dynamic power spectrum generated from the GX-339 light curve shown in Figure \ref{fig:psd}. 
We generated 63 light curve segments of 16 seconds length with a $0.02\,\mathrm{s}$ time resolution and Fourier-transformed each to generate a power spectrum. 
The dynamic power spectrum here plots each power spectrum as a vertical slice as a function of time, with the color indicating the fractional rms-squared-normalized power in each bin (yellow are large powers; purple, small) . 
The dynamic power spectrum was clipped to around the range of the QPO, and smoothed using bicubic interpolation to improve clarity. 
The QPO is clearly visible as a yellow streak, and seems not to be present during the entire observation \citep[consistent with][]{belloni2005}. 
In red, we show the frequency with the highest power found in each segment (excluding frequencies below $3\,\mathrm{Hz}$ to exclude the low-frequency red noise), using the \texttt{trace\_maximum} method.}
\label{fig:dynspec}
\end{center}
\end{figure}

For long observations with quasi-periodic oscillations (QPOs) spectrograms, more commonly known in the astronomy literature as Dynamic Power Spectra, can be a useful way to track changes in the QPO centroid frequency over time. 
We have implemented \texttt{DynamicalPowerspectrum} as a subclass to \texttt{AveragedPowerspectrum} to provide this functionality. 
Like \texttt{AveragedPowerspectrum}, this class takes a \lightcurve object and a segment size as input, but instead of averaging the power spectra of each individual segment, it will create a matrix of time bins (one bin for each segment) as columns and Fourier frequencies as rows. 
Rebinning both along the time and frequency axis is possible. 
Moreover, the method \texttt{trace\_maximum} automatically finds the frequency with the highest power in each segment in a given range of frequencies, and traces this maximum over time. 
An example using data from the source GX~339--4 is shown in Figure \ref{fig:dynspec}.

Closely related to the cross spectrum and power spectrum are the crosscorrelation and the autocorrelation, implemented in classes \texttt{CrossCorrelation} and \texttt{AutoCorrelation}. 
As their respective Fourier spectra equivalents they take either one (autocorrelation) or two (cross correlation) \lightcurve objects as input and computes the correlation between the two light curves or of the single light curve with itself, along with the time lags for which the correlation was produced and the time lag at which the maximum correlation is measured.

It is useful to note that all classes in this section are compatible with GTIs. 
The classes \powerspectrum and \crossspectrum will generate warnings if the observations contain gaps; their averaged versions will take GTIs correctly into account by producing power spectra only from light curve segments for which data is available.  


\section{The \texttt{modeling} Subpackage}
\label{sec:modeling}

Modeling data sets with parametric (often physically motivated) models that map an independent variable (e.g., time or frequency) to one or more dependent variables (e.g., flux, counts or Fourier powers) is a common task in astronomy. 
Constructing a universal modeling framework is a highly non-trivial task, and excellent packages exist for general-purpose model building (e.g., STAN, \citealt{stan}). 
Thus, \stingray's modeling interface restricts itself to models of commonly used spectral-timing products, in particular (averaged) power spectra. 
While it makes heavy use of the \verb|astropy.modeling.FittableModel| definitions, it uses custom definitions for fitting algorithms motivated by the statistical properties of spectral timing products, which deviate significantly from other data types commonly found in astronomy and thus cannot easily be modelled with standard approaches defined in \texttt{astropy}.

The modeling subpackage logically separates out statistical models -- likelihoods and posteriors -- from the fitting functionality, such that different likelihoods and posteriors can be straightforwardly dropped in and out depending on the data set and problem at hand. 
In line with the overall philosophy of \stingray, the modeling subpackage is designed to be modular and easily extensible to specific problems a user might try to solve, while many typical tasks one might do with Fourier products are already built-in. 
It makes use of the \verb|scipy.optimize| interface for optimization as well as the package \texttt{emcee} for Markov Chain Monte Carlo (MCMC) sampling.

\subsection{Statistical Models}

All statistical models are implemented as a subclass of an Abstract Base Class \likelihood in module \verb|stingray.posterior|. 
In its most basic form, each subclass of \likelihood takes data in some form (most commonly two arrays, one with the independent and one with the dependent variable) as well as an object of type \verb|astropy.modeling.FittableModel|. 
The likelihood computes model values for each data point in the array of independent variables and statistically compares these model values with the data points stored in the dependent variable, assuming the particular statistical distribution of the likelihood definition. 
The result is a single scalar, which can then be, for example, used in an optimization algorithm in order to find a Maximum Likelihood (ML) solution.

For all likelihoods in \stingray, an equivalent subclass of \verb|stingray.modeling.Posterior| is available, which uses the \likelihood definitions to compute posterior probability densities for the parameters of a model given data.
All subclasses of \verb|Posterior| also require definition of a \texttt{logprior} method, which calculates the value of the prior probability density of the parameters. 
Because priors are strongly problem-dependent, they cannot be hard-coded into \stingray. 
Even for relatively straightforward problems such as modeling quasi-periodic oscillations of X-ray binaries, the physical properties and their effect on the data can differ strongly from source to source, indicating that a prior set for XTE~J1550--564 may not be appropriate for e.g.~GRS~1915+105. 
Separating out the likelihood and posterior in distinct classes makes it possible to allow the use of the likelihood for maximum likelihood estimation, while requiring priors for estimating the Bayesian posterior probability through, e.g., MCMC simulations.

\texttt{Loglikelihood} and \texttt{Posterior} subclass definitions currently exist within \stingray for different statistical models useful in the context of astronomical data. 
\verb|GaussianLogLikelihood| and \texttt{GaussianPosterior} implement statistical models for data with normally distributed uncertainties. \texttt{GaussianLogLikelihood} will compute what astronomers generally call $\chi^2$, because the likelihood calculated by this statistical model generally follows a $\chi^2$ distribution with $N-P$ degrees of freedom (where $N$ is the number of data points and $P$ the number of free parameters). Note, however, that this is \textbf{not} the same as the $\chi^2$ likelihood defined below!

\texttt{PoissonLogLikelihood} and \texttt{PoissonPosterior} calculate the likelihood and posterior for Poisson-distributed data, respectively. This likelihood is equivalent to what in astronomy is often called the \textit{Cash statistic} \citep{cash1979} and is the appropriate likelihood to use for count- or event-type data often found in X-ray astronomy time series and spectra.

\texttt{PSDLogLikelihood} and \texttt{PSDPosterior} implement the statistical model appropriate for modeling (averaged) power spectra, a $\chi^2$ distribution. 
We broke with the rule of naming likelihoods and posteriors after the statistical distribution they implement in this case, because as mentioned above, astronomers tend to call the likelihood for normally distributed data $\chi^2$, and this naming helps avoid any confusion. 
These two classes implement a $\chi^2_2$ distribution for Fourier spectra generated with the \powerspectrum class, and a $\chi^2_{2MK}$ distribution for power spectra generated with the \texttt{AveragedPowerspectrum} class, where $M$ is the number of averaged segments and $K$ is the number of averaged neighbouring frequency bins. 
Please note that as laid out in \citet{huppenkothen2017}, these distribution are \textbf{not} appropriate for use on (averaged) cross spectra. 
The appropriate distributions for these products are under development for the next version of \stingray.

Other statistical models can be easily implemented by subclassing the \texttt{LogLikelihood} and \texttt{Posterior} Abstract Base Classes and using the existing classes as template.

\begin{figure*}[htbp]
\begin{center}
\subfloat[]{%
\includegraphics[width=10cm]{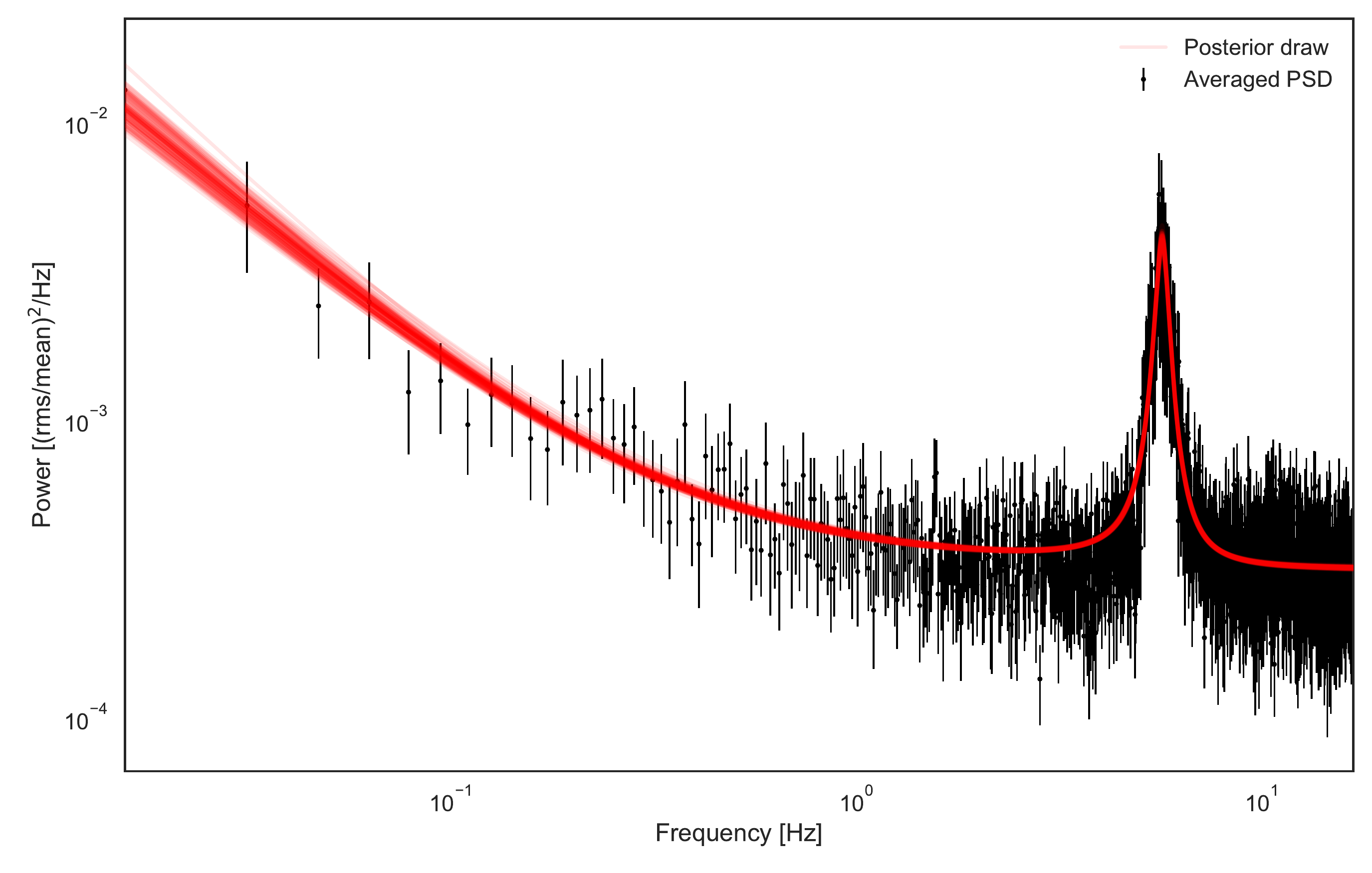}
}\subfloat[]{%
\includegraphics[width=6.5cm]{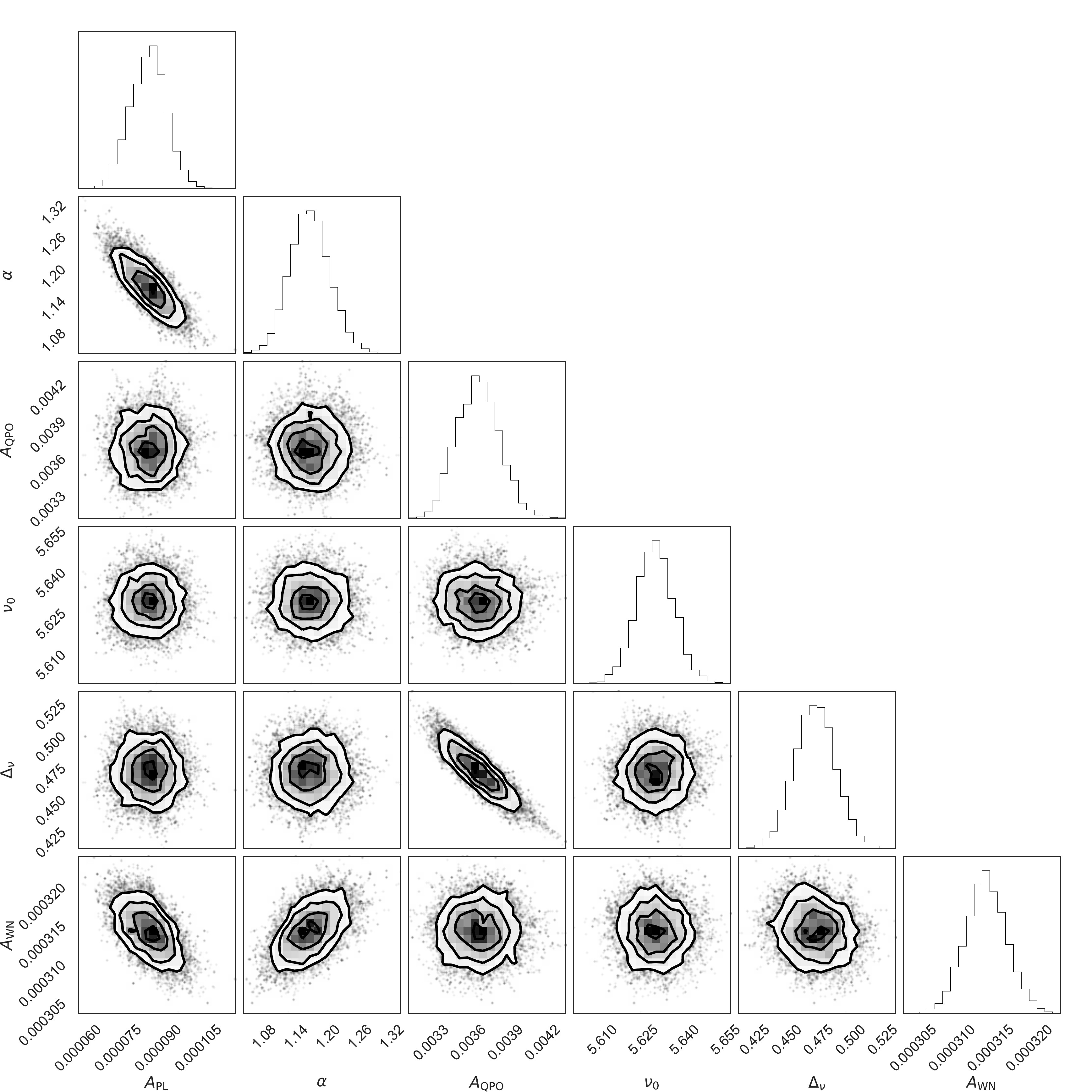}
}\caption{Left panel: In black, a power spectrum averaged out of 15 light curve segments of 64s each of the GX~339--4 observation, along with draws from the posterior distribution of the power law model plus the Lorentzian QPO model and constant used to represent the data (red). 
Right panel: corner plot showing the marginal posterior distributions (diagonal) of the six parameters of the model: the amplitude of the power law $A_\mathrm{PL}$, the power law index $\alpha$, the amplitude of the Lorentzian $A_\mathrm{QPO}$, the QPO centroid frequency $\nu_0$, the width of the QPO $\Delta_\nu$, and the amplitude of the white noise, $A_\mathrm{WN}$. The right-hand figure was produced using the package \texttt{corner} \citep{corner}.}
\label{fig:posterior}
\end{center}
\end{figure*}

\subsection{General Parameter Estimation and Model Comparison Functionality}

\stingray implements utility functions in order to reduce some of the overhead required for standard parameter estimation and model comparison tasks. 
In particular, the \verb|parameterestimation| module implements classes and functions to aid users in fitting models to data and estimating the probability distributions of parameters.

The class \texttt{ParameterEstimation} provides the basis for more sophisticated, specialized implementations for particular data types. 
Its core methods are \verb|fit| and \verb|sample|. 
The former takes an instance of a \verb|LogLikelihood| or \verb|Posterior| subclass and uses minimization algorithms implemented in \verb|scipy.optimize| to find the Maximum Likelihood (ML) or Maximum-A-Posteriori (MAP) solution. The \verb|sample| method uses the Affine-Invariant MCMC sampler implemented in \texttt{emcee} \citep{emcee} to generate samples from a posterior distribution passed as an instance of a subclass of \verb|Posterior|. Note that \textit{you should never pass a} \verb|LogLikelihood| \textit{instance into the} \verb|sample| \textit{method}, because sampling from a likelihood is statistically invalid. In addition to these core methods, higher-level functionality implemented in this class includes calculating the Likelihood Ratio Test (LRT) for two different models $M_1$ and $M_2$ via the \verb|compute_lrt| method (note the statistical assumptions of the LRT, and where they fail, e.g., \citealt{protassov2002}). In addition, the \verb|calibrate_lrt| method allows calibrating the p-value for rejecting the model $M_1$ via simulations of $M_1$, using either an MCMC sample (for Bayesian inference and posterior predictive $p$-values) or the covariance matrix derived from the optimization (both Bayesian and Maximum Likelihood approaches).

\stingray also implements two classes that summarize results of the optimization and sampling procedures in concise, useful ways. 
The \verb|fit| method returns an instance of class \verb|OptimizationResults|. 
This contains the most important outputs from the optimizer, but will also behind the scenes calculate a number of useful quantities, including the covariance between parameters (or a numerical approximation for some minimization algorithms), the Akaike and Bayesian Information Criteria (AIC: \citealt{akaike1974}; BIC: \citealt{schwarz1978}) as well as various summary statistics.

Similarly, an instance of class \verb|SamplingResults| is returned by the \verb|sample| method, which returns the posterior samples calculated by the MCMC sampler, as well as computes a number of helpful quantities using the MCMC chains. It calculates useful diagnostics including the acceptance fraction, the autocorrelation length and the Rubin-Gelman statistic \citep{gelman1992} to indicate convergence, and infers means, standard deviations and user-defined credible intervals for each parameter.

\subsection{Special Functionality for Fourier Products}

The subclass \verb|PSDParEst| implements a number of additional methods particularly useful for modeling power spectra. 
One particularly common task is to search for periodic signals (e.g., from pulsars) in a power spectrum, which reduces to finding outliers around an assumed power spectral shape (assuming the signal is \textit{strictly} periodic, and thus all power approximately concentrated in one bin). 
In the presence of other variability, the probability of observing a certain power $P_j$ at a frequency $\nu_j$ under the assumption that no periodic signal is present depends on the shape and parameters of the underlying power spectral model assumed to have generated the data. 
As \citet{vaughan2010} show, there is an inherent uncertainty in our inference of the parameters of this power spectral model, which must be taken into account via simulations. 
\verb|PSDParEst| implements a method \verb|calibrate_highest_outlier|, which finds the $k$ highest outliers (where $k$ is a user-defined number) and calculates the posterior predictive $p$-value that said outliers cannot be explained by noise alone. 
It makes heavy use of the method \verb|simulate_highest_outlier|, which uses the \verb|sample| method to derive an MCMC sample and then simulate fake power spectra from that model for a range of plausible parameter values in order to include our model uncertainty in the posterior predictive $p$-value. 
For details of the overall procedure, see \citet{vaughan2010}. 

As of this version, the \verb|stingray.modeling| subpackage has no functionality to model higher-order Fourier products. 
For spectral timing in particular, this would involve being able to read and apply instrument responses to models, as well as being able to interface with the library of spectral models associated with the X-ray spectral fitting tool XSPEC \citep{arnaud1996}. 
Providing this functionality is planned for a future release of \stingray. 

\begin{figure*}[htbp]
\begin{center}
\includegraphics[width=\linewidth]{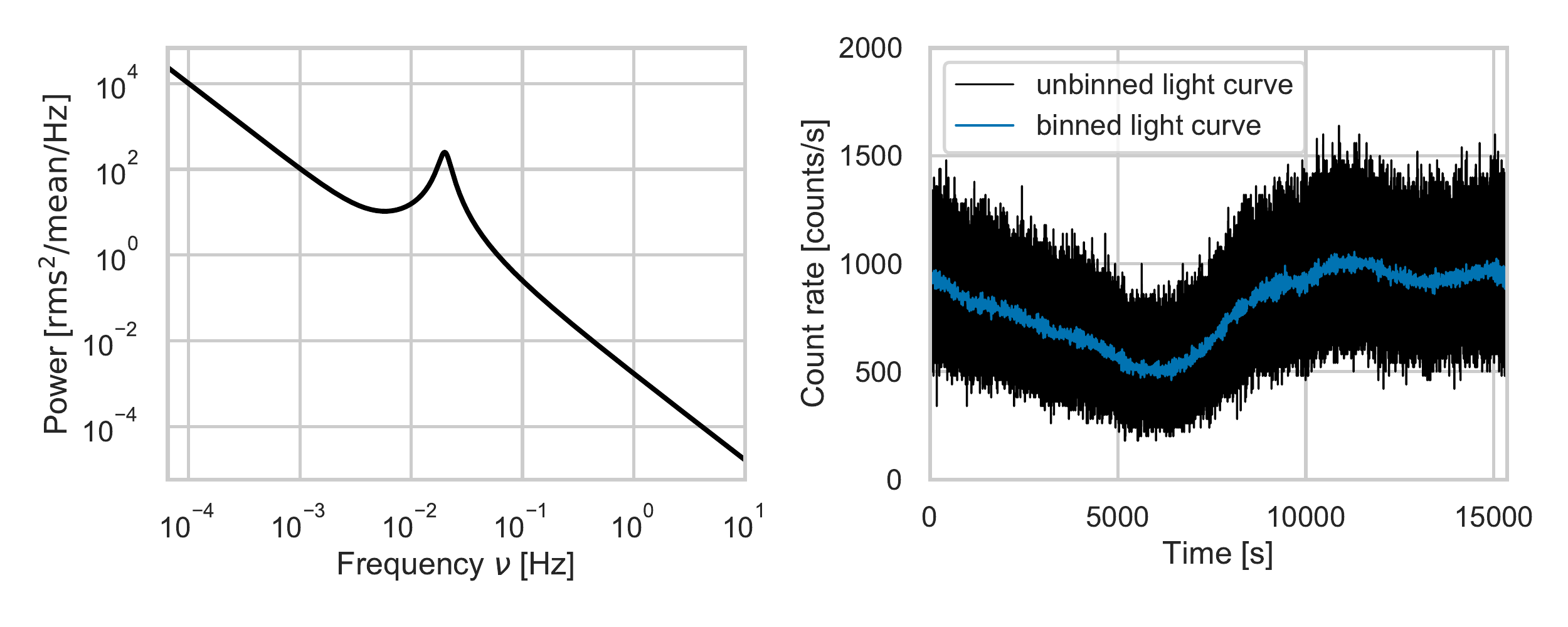}
\caption{Left: A power spectral shape generated using a compound \texttt{astropy.modeling.models} object of a power law and a Lorentzian. 
Right: A corresponding light curve generated by the \texttt{simulator} subpackage with a time resolution of $0.05\,\mathrm{s}$, a total duration of $15\,\mathrm{ks}$, a mean count rate of $40\,\mathrm{counts}/\mathrm{s}$ and a fractional rms amplitude of $0.2$. In blue we show a binned version of the same light curve.}
\label{fig:sim_lc}
\end{center}
\end{figure*}

\section{The \texttt{simulator} Subpackage}
\label{sec:simulator}

The \texttt{simulator} subpackage contains a number of methods to generate simulated light curves out of known power spectral shapes, and event lists from light curves. 

\subsection{Simulating light curves from input power spectra}
The basic \texttt{Simulator} object uses the algorithm from \citet{timmer1995} to generate light curves out of a spectral shape. 
The spectral shape can be input as a spectral power-law index, \texttt{astropy.modeling.models} objects, as well as a user-given array of powers.
In Figure \ref{fig:sim_lc}, we present a light curve as generated by a given power spectral shape. 
The output is a \texttt{Lightcurve} object that can be used like real data sets, including all functionality related to GTIs, spectral-timing products and modeling. 

\subsection{Use transfer functions on light curves}
Most astrophysical signals we receive in our instruments are the mixture of different input signals. 
Often, a signal emitted in one region can be reflected and re-emitted from another region, or filtered in different ways. 
Spectral timing studies can decompose the signals and try to understand how the signal is transformed by these phenomena between the emission region and the observer \citep[see][for a review]{uttley2014}.
This transformation can be encoded in an impulse response function, that describes the response of the system to a delta-function impulse. 
This is the Fourier transform of another well-known quantity in signal processing, the transfer function \citep[see][]{girod2001signals}.
The \texttt{Simulator} object is capable of generating multiple light curves starting from an initial light curve and multiple input responses, mimicking observations in different energy bands. 

\subsection{Simulating event lists from light curves}
The \texttt{simulator.base.simulate\_times} method is able to simulate event lists from input light curves.
It implements the acceptance-rejection method: 
\begin{enumerate}
\item Generate a light curve (and smooth out any Poisson noise if generated through the \citealt{timmer1995} method) over the whole observation; normalize it so that the maximum is 1; 
\item Generate an event, with uniform probability over the observing time; 
\item Associate to this event a uniform random number $\mathcal{P}$ between 0 and 1; 
\item If $\mathcal{P}$ is lower than the normalized light curve at the event time, \textit{accept} the event, otherwise \textit{reject} it.
\end{enumerate}
In \stingray, we use arrays of events for better performance, using the functionality contained in the \texttt{numpy} library.


\section{The \texttt{pulse} Subpackage}
\label{sec:pulsar}
The subpackage \texttt{pulse} contains the basic operations to perform the search and characterization of pulsed signals for use e.g.\ in searches of X-ray pulsars.

\subsection{Epoch Folding}
Among the basic algorithms used in pulsar astronomy, one cannot overstate the importance of Epoch Folding (EF).
The algorithm consists of cutting the signal at every pulse period and summing all sub-intervals in phase. 
An alternative way of seeing it, more useful for photon data, is as a \textit {histogram of pulse phases}.

If the period is exactly correct and assuming a stable pulsation, the signal-to-noise ratio will get better approximately with the square root of the number of summed sub-intervals.
This is the method used to obtain practically all pulse profiles shown in the literature, as most pulsar signals are orders of magnitude below the noise level.

The \texttt{pulse.pulsar} submodule contains the functionality to calculate the phase given a simple pulse ephemeris consisting of any number of pulse frequency derivatives, or using a number of methods for the orbit of the pulsar (using the optional dependency \texttt{PINT}).
Moreover, the module also includes a mechanism to calculate the exposure of single bins in the pulse profile. 
This is particularly useful for very long-period pulsars where the pulsed period is comparable to the length of the GTIs.
The different exposure of pulse bins caused by the absence of signals during GTIs is taken into account in the calculation of the final pulse profile by the folding algorithm, if the user asks for it. 

\subsection{Epoch Folding Searches and \zsq Searches}
\label{sec:efzsq}
During a search for pulsations, the first step is usually calculating a power spectrum through a Fast Fourier Transform. 
However, often pulsations do not leave a clear signature above the noise level in the power spectrum, because they are weak or they fall close to bin edges, where the sensitivity is reduced.\footnote{This is due to the convolution of the signal with the observing window, that produces a sinc-like response inside the bins of the FFT; periodic signals with the same amplitude are detected with a lower Fourier amplitude if they fall far from the center of the spectral bin \citep{vanderklis1989}.}
Even when the signature is clear, the frequency resolution of the power spectrum is often inadequate to measure precisely the pulse frequency.
Therefore, an additional statistical analysis is needed. 

\stingray implements two statistical methods for pulsar searches, that can be applied to event lists or light curves (that are treated as event lists with ``weights'').

The Epoch Folding Search (EFS) method consists of executing the folding at many trial frequencies around the candidate frequency.
Once the folding is performed, the following statistics is calculated on the profile:
\begin{equation}
\mathcal{S} = \sum_i\frac{(P_i - \overline{P})^2}{\sigma^2}
\end{equation}
where $P_i$ are the bins of the profile, $\overline{P}$ is the mean level of the profile, and $\sigma$ is the standard deviation.
$\mathcal{S}$ is the summed squared error of the actual pulsed profile with respect to a \textit{flat} model, and follows a $\chi^2$ distribution.

If there is no pulsation, $\mathcal{S}$ will assume a random value distributed around the number of degrees of freedom $n - 1$ (where $n$ is the number of bins in the profile) with a well defined statistical distribution ($\chi^2_{n - 1}$) that allows an easy calculation of detection limits. 
When observing a peak of given height is very unlikely under the null hypothesis (meaning that the probability to obtained this peak by noise is below a certain $\epsilon$), this peak is considered a pulse candidate.
If the frequency resolution is sufficiently high, close to the correct frequency, as described by \citet{leahy1983b} and \citet{leahy1987}, the peak in the epoch folding periodogram has the shape of a $\mathrm{sinc}^2$ function whose width is driven by the length $T$ of the observation (FWHM $\Delta \nu\sim0.9/T$).

The epoch folding statistic, however, can give the same value for a pulse profile at the correct frequency and, for example, a harmonic that produces a deviation from a Poisson distribution.
A more effective method is the $Z^2_n$ statistics \citep{buccheri1983}, which is conceptually similar to EF but has high values when the signal is well described by a small number of sinusoidal harmonics: 

\begin{equation}
\zsq = \dfrac{2}{N} \sum_{k=1}^n \left[{\left(\sum_{j=1}^N \cos k \phi_j\right)}^2 + {\left(\sum_{j=1}^N \sin k \phi_j\right)}^2\right] \; ,
\end{equation}

\noindent where $N$ is the number of photons, $n$ is the number of harmonics, $\phi_j$ are the phases corresponding to the event arrival times $t_j$ ($\phi_j = \nu t_j$, where $\nu$ is the pulse frequency).

The \zsq statistics defined in this way, far from the pulsed profile, follows a $\chi^2_n$ distribution, where $n$ is the number of harmonics this time.
This allows, again, to easily calculate thresholds based on the probability of obtaining a given \zsq by pure noise.

The standard \zsq search calculates the phase of each photon and calculates the sinusoidal functions above for each photon.
This is very computationally expensive if the number of photons is high. 
Therefore, in \stingray, the search is performed by binning the pulse profile first and using the phases of the folded profile in the formula above, multiplying the squared sinusoids of the phases of the pulse profile by a weight corresponding to the number of photons at each phase.

\begin{equation}
\zsq \approx \dfrac{2}{\sum_j{w_j}} \sum_{k=1}^n \left[{\left(\sum_{j=1}^m w_j \cos k \phi_j\right)}^2 + {\left(\sum_{j=1}^m w_j \sin k \phi_j\right)}^2\right]
\end{equation}

Since the sinusoids are only executed on a small number of bins, while the epoch folding procedure just consists of a very fast histogram-like operation, the speedup of this new formula is obvious. 
Care must be put into the choice of the number of bins, in order to maintain a good approximation even when the number of harmonics is high. 
We recommend in the documentation to use a number of bins at least 10 times larger than the number of harmonics.

\begin{figure}[htbp]
\begin{center}
\includegraphics[width=\linewidth]{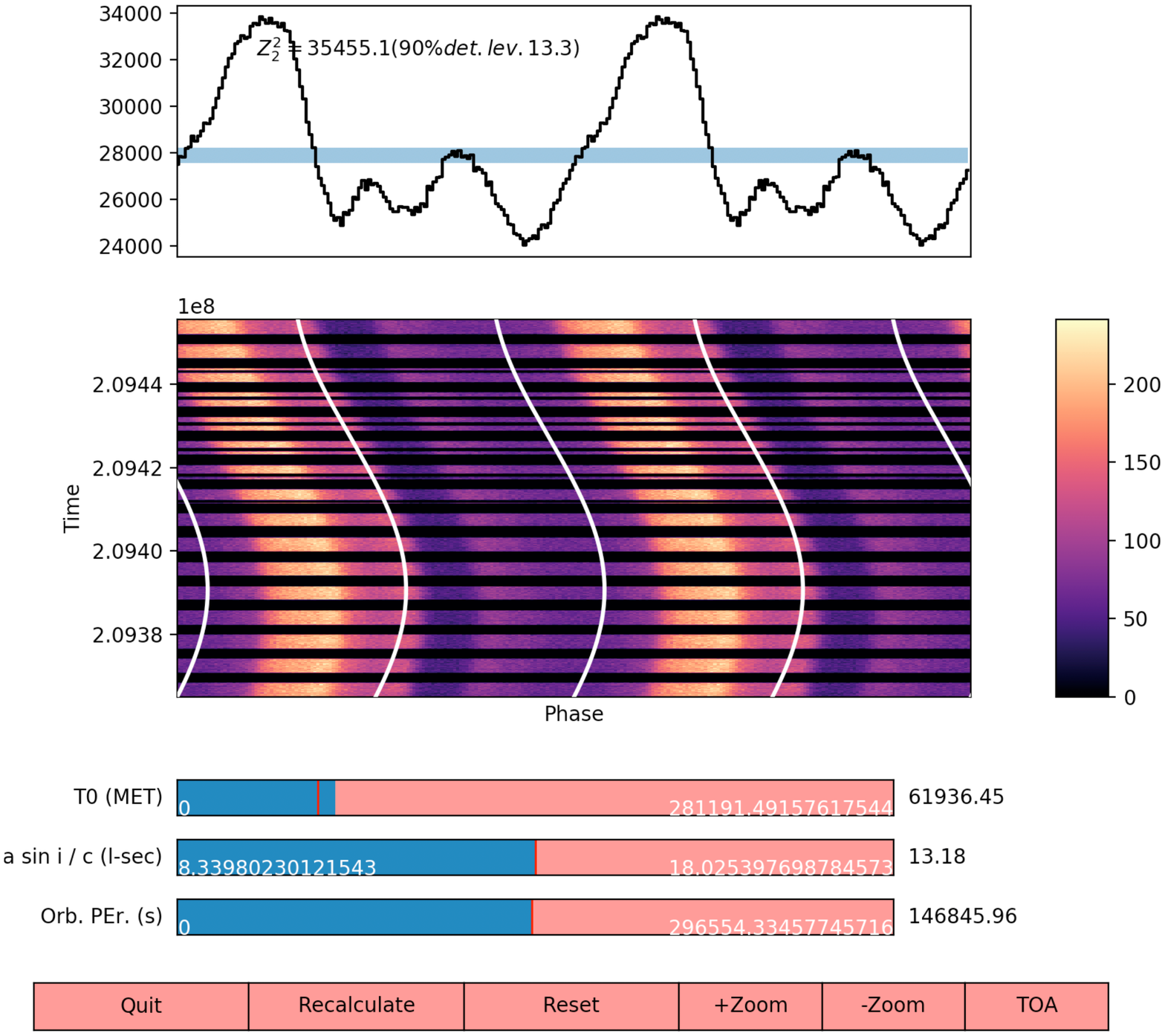}
\caption{Phaseogram showing the variation of the pulse phase corresponding to an imperfect orbital solution (in this case the time at the ascending node $T_0$) in a \nustar observation of Her X-1, executed with \stingray and plotted in a convenient, interactive interface with \hendrics. 
The TOA button allows the user to calculate the TOA for use with Tempo2, PINT or similar programs.}
\label{fig:phaseogram}
\end{center}
\end{figure}

\subsection{Characterization of pulsar behavior}
\label{sec:ephem}
As seen in Section~\ref{sec:efzsq}, the \zsq or the EF periodograms of a perfectly stable pulsation have the shape of a $\mathrm{sinc}^2$ function.
\stingray has functionality to fit these periodograms with a $\mathrm{sinc}^2$ function or alternatively a Gaussian model, and find the mean frequency with high precision\footnote{When using the Gaussian model, the width of the impulse is similar to the  FWHM $\Delta \nu\sim0.9/T$ of the  $\mathrm{sinc}^2$ function (Section \ref{sec:efzsq}). It does \textit{not} represent an errorbar to the frequency measurement.}.

A significant deviation from the expected shape from these models can happen if the pulsation is not stable.
Calculating the \textit{phaseogram} (Figure~\ref{fig:phaseogram}) is an option to investigate how the pulse phase varies in time.
The phaseogram in this context consists of a 2D histogram of the phase and arrival times of the pulses. 
If the pulsation is stable and the pulse frequency was determined with precision, the phaseogram shows vertical stripes corresponding to perfectly aligned pulses.
If the frequency is not as precise, the stripes become more and more diagonal.
If the pulse has a detectable frequency derivative, these stripes bend with a parabolic shape.
If the orbital solution is imperfect, the stripes show specific periodic features\footnote{See for example \url{https://github.com/matteobachetti/timing-lectures/blob/master/no-binder/Timing_residuals.ipynb}}.

A very precise way to determine the exact pulse ephemeris is out of the scope of \stingray. 
Nonetheless, \stingray has a mechanism to calculate the pulse arrival times (or times of arrival, TOAs) to be analyzed with more specialized software like Tempo, Tempo2 or PINT. 
We use the same \texttt{fftfit} algorithm used for radio pulsars \citep{Taylor92}, that calculates the cross-correlation between a template profile and the folded profile in the Fourier domain. 
This is implemented in the \texttt{get\_TOA} function in \texttt{stingray.pulse.pulsar}.

The functionality to plot the phaseogram, interactively change the timing parameters (either pulse parameters or orbital parameters) and adjusting the solution, and calculating the TOAs for use with external programs, is conveniently accessible in \hendrics (See Section \ref{sec:hendrics}) and Figure~\ref{fig:phaseogram} and in \dave (Section \ref{sec:dave}).

\section{\hendrics: A Command-Line Interface for \stingray}
\label{sec:hendrics}

The \hendrics package\footnote{\url{https://github.com/stingraySoftware/hendrics}}---formerly called \texttt{MaLTPyNT} \citep{bachetti2015b}---builds upon \stingray by providing a suite of easy-to-execute command-line scripts whose primary use is providing an accurate quick-look (spectral-)timing analysis of X-ray observations, useful for a range of use cases, including exploratory data analysis and quality assessment of larger data analysis pipelines. 
While its initial development proceeded independently from \stingray, much of its core functionality since version 3.0 is based on the classes and methods \stingray provides, and some key functionality has been shifted to \stingray where appropriate. 

Its key distinguishing feature from established command-line interfaces such as FTOOLS is the accurate treatment of gaps in the data (for example due to the Earth's occultation or the South Atlantic Anomaly), as well as its treatment of dead time for certain detectors like \nustar. 
Where \stingray aims to provide flexible building blocks for designing sophisticated spectral-timing analysis workflows, \hendrics provides end-to-end solution for common tasks such as power- and cross spectra, time lags, pulsar searches, color-color as well as color-intensity diagrams, at the cost of loosing some flexibility during the creation of those products. 
Like \stingray, \hendrics is an \astropy affiliated package and aims to build upon and be compatible with functionality provided as part of the \astropy ecosystem.
\hendrics supports a range of output data formats including netCDF4 and ASCII formats, which can then be read into other astronomical data analysis systems such as XSPEC \citep{arnaud1996} or ISIS  \citep{houck2000}.

\hendrics is in release version 4 as of 2018-02-12, and under active development, utilizing the same continuous integration, testing and code review standards as \stingray.

\begin{figure*}[htbp]
\begin{center}
\includegraphics[width=\linewidth]{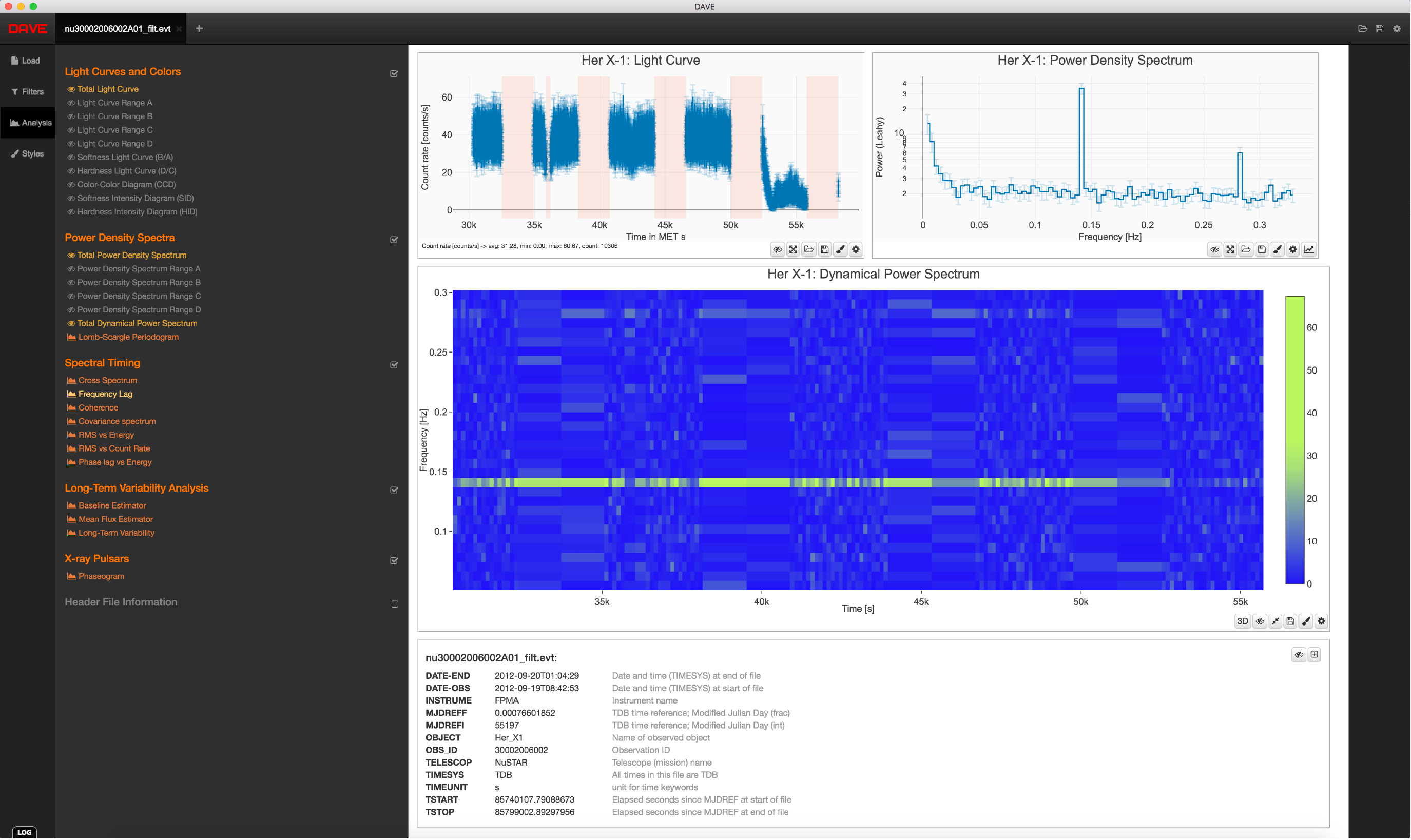}
\caption{An example of the \dave graphical user interface for the Her X-1 pulsar data observed with \nustar: 
In the top left, we show the last 30\,ks of the pulsar light curve with the GTIs clearly marked. In the top right, we plot the averaged power spectrum generated from 109 segments of $256\,\mathrm{s}$ duration with a binned time resolution of $1.5\,\mathrm{s}$. 
In the middle, we present the dynamic power spectrum generated from the same $256\,\mathrm{s}$ segments that generated the top right averaged power spectrum. 
Below, header meta-data is shown for reference. 
On the left, the menu presents a range of options of figures to plot and compare, including spectral-timing capabilities. 
All figures are interactive, including panning and zooming, as well as interactive choices of data selection.}
\label{fig:dave}
\end{center}
\end{figure*}

\section{\texttt{DAVE}: Exploratory Data Analysis in a Graphical User Interface}
\label{sec:dave}

\dave\footnote{\url{https://github.com/stingraySoftware/dave}}---the Data Analysis for Variable Events package---is a Graphical User Interface built on top of \stingray in order to provide users with interactive capabilities for exploratory data analysis of variable time series. 
Much of the core functionality within \stingray is available in \dave as well: creation of power spectra, cross spectra, dynamical power spectra and spectral-timing products such as time lags and coherence measurements. 
In addition, it implements interactive filtering of light curves with respect to energy channels or energies (if a response matrix file is loaded), time ranges, and count rates. 
Users may compare light curves and power spectra from different energy ranges, and may create auxiliary products such as color-color and color-intensity diagrams that further aid the exploration of the data. 
An example of the interface is shown in Figure \ref{fig:dave}. 
The full interface and its capabilities will be described in a future publication.


\section{Future Development Plans}
\label{sec:future}

Near- and medium-term plans for \stingray development are largely aimed at extending current functionality related to Fourier spectra, and continuing work towards comprehensive spectral-timing capabilities. 
While open-source reference implementations of higher-order Fourier products such as bispectra, biphase and bicoherence exist \citep{maccarone2002,maccarone2005,maccarone2013}, they require additional extensions to be useful for X-ray spectral timing. 
New key features in the next version of \stingray, based on an existing reference implementation of covariance spectra \citep{WilkinsonUttley09}, will include lag-energy spectra \citep{Vaughanetal94}, rms-energy spectra \citep{Revnivtsevetal99}, and excess variance spectra \citep{Vaughanetal03}. 
In addition, while at the moment there is rudimentary functionality to build spectra-timing products, it is currently not possible to seamlessly work with these products using \stingray, because \stingray currently has no native capability for energy-spectral modeling. 
Instead, they would have to be exported (e.g., saved to disk) and then loaded into another software, significantly disrupting workflows and pipeline development. 
In order to streamline this process, we aim to connect \stingray with existing packages for modeling X-ray spectra. 
Here, it will be necessary to connect \stingray with the extensive suite of physical models implemented in XSPEC, as well as existing spectral fitting codes implemented in \textit{Python}, most notably the open-source package \textit{Sherpa} \citep{sherpa}.

Data rates from current and future X-ray instruments are increasing at a precipitous rate, pushing memory and processing requirements for even simple tasks like Fast Fourier Transforms of data observed e.g. with \textit{NICER} and \textit{Astrosat} into a regime that is difficult with standard desktop computing architectures. 
Therefore, the other strong emphasis for the second version of \stingray will be code and algorithm optimization. 
Where possible, we will replace existing implementations by high-performance equivalents that take advantage of recent developments in computing (such as GPU-enabled computations and multi-core batch processing), optimize and streamline existing code to minimize computational overhead and memory usage of the classes and functions implemented within \stingray. 

Typical X-ray timing observations--long, deep stares at single objects with high time resolution--are particularly well suited for Fourier-based methods, owing to their relatively regular time sampling and observation duration much longer than the physical time scales of interest encoded in the light curves. While some optical telescopes, most notably \textit{Kepler} and the Transiting Exoplanet Survey Satellite (\textit{TESS}; \citealt{ricker2015}), employ similar modes of observation that make the current suite of methods in \stingray transferable to data from these instruments, many current and future \textit{survey instruments} such as the \textit{Zwicky Transient Facility} (\textit{ZTF}; \citealt{bellm2019,graham2019}) or the \textit{Large Synoptic Survey Telescope} (\textit{LSST}; \citealt{lsst2019}) will provide the community with long-baseline light curves that are irregularly sampled. A range of methods has been developed for time series analysis of these light curves in the time domain, e.g.\ CARMA (see for example \citealt{kelly2014,foremanmackey2017}), and ARIMA \citep[e.g.][]{feigelson2018} processes. Developing \textit{spectral} timing methods for irregularly sampled light curves is a major future challenge for the field, and a high-priority long-term goal for \stingray.


\acknowledgments

We thank Astro Hack Week for providing the venue that started this project and the Lorentz Center workshop `The X-ray Spectral-Timing Revolution' (February 2016) that started the collaboration, as well as the Google Summer of Code Program for funding a total of 5 co-authors (OM, UMK, HM, HR, SS) who implemented a large fraction of the various library components over three summers. We thank Jim Davenport for providing help with the Kepler data. We are grateful to Matteo Guanazzi and Erik Kuulkers for their advice and suggestions for implementing DAVE, as well as the anonymous referee and the AAS statistics editor for their helpful comments on the manuscript.
D.H. acknowledges support from the DIRAC Institute in the Department of Astronomy at the University of Washington. 
The DIRAC Institute is supported through generous gifts from the Charles and Lisa Simonyi Fund for Arts and Sciences, and the Washington Research Foundation.
M.B. is supported in part by the Italian Space Agency through agreement ASI-INAF n.2017-12-H.0 and ASI-INFN agreement n.2017-13-H.0.
A.L.S. is supported by an NSF Astronomy and Astrophysics Postdoctoral Fellowship under award AST-1801792.
E.M.R. acknowledges support from Conselho Nacional de Desenvolvimento Cient\'ifico e Tecnol\'ogico (CNPq -- Brazil)






\bibliography{stingraypaper}

\begin{thebibliography}{}
\expandafter\ifx\csname natexlab\endcsname\relax\def\natexlab#1{#1}\fi
\providecommand{\url}[1]{\href{#1}{#1}}
\providecommand{\dodoi}[1]{doi:~\href{http://doi.org/#1}{\nolinkurl{#1}}}
\providecommand{\doeprint}[1]{\href{http://ascl.net/#1}{\nolinkurl{http://ascl.net/#1}}}
\providecommand{\doarXiv}[1]{\href{https://arxiv.org/abs/#1}{\nolinkurl{https://arxiv.org/abs/#1}}}

\bibitem[{{Abdo} {et~al.}(2013){Abdo}, {Ajello}, {Allafort}, {Baldini},
  {Ballet}, {Barbiellini}, {Baring}, {Bastieri}, {Belfiore}, {Bellazzini},
  {Bhattacharyya}, {Bissaldi}, {Bloom}, {Bonamente}, {Bottacini}, {Brandt},
  {Bregeon}, {Brigida}, {Bruel}, {Buehler}, {Burgay}, {Burnett}, {Busetto},
  {Buson}, {Caliandro}, {Cameron}, {Camilo}, {Caraveo}, {Casandjian}, {Cecchi},
  {{\c{C}}elik}, {Charles}, {Chaty}, {Chaves}, {Chekhtman}, {Chen}, {Chiang},
  {Chiaro}, {Ciprini}, {Claus}, {Cognard}, {Cohen-Tanugi}, {Cominsky},
  {Conrad}, {Cutini}, {D'Ammando}, {de Angelis}, {DeCesar}, {De Luca}, {den
  Hartog}, {de Palma}, {Dermer}, {Desvignes}, {Digel}, {Di Venere}, {Drell},
  {Drlica-Wagner}, {Dubois}, {Dumora}, {Espinoza}, {Falletti}, {Favuzzi},
  {Ferrara}, {Focke}, {Franckowiak}, {Freire}, {Funk}, {Fusco}, {Gargano},
  {Gasparrini}, {Germani}, {Giglietto}, {Giommi}, {Giordano}, {Giroletti},
  {Glanzman}, {Godfrey}, {Gotthelf}, {Grenier}, {Grondin}, {Grove},
  {Guillemot}, {Guiriec}, {Hadasch}, {Hanabata}, {Harding}, {Hayashida},
  {Hays}, {Hessels}, {Hewitt}, {Hill}, {Horan}, {Hou}, {Hughes}, {Jackson},
  {Janssen}, {Jogler}, {J{\'o}hannesson}, {Johnson}, {Johnson}, {Johnson},
  {Johnson}, {Johnston}, {Kamae}, {Kataoka}, {Keith}, {Kerr}, {Kn{\"o}dlseder},
  {Kramer}, {Kuss}, {Lande}, {Larsson}, {Latronico}, {Lemoine-Goumard},
  {Longo}, {Loparco}, {Lovellette}, {Lubrano}, {Lyne}, {Manchester}, {Marelli},
  {Massaro}, {Mayer}, {Mazziotta}, {McEnery}, {McLaughlin}, {Mehault},
  {Michelson}, {Mignani}, {Mitthumsiri}, {Mizuno}, {Moiseev}, {Monzani},
  {Morselli}, {Moskalenko}, {Murgia}, {Nakamori}, {Nemmen}, {Nuss}, {Ohno},
  {Ohsugi}, {Orienti}, {Orlando}, {Ormes}, {Paneque}, {Panetta}, {Parent},
  {Perkins}, {Pesce-Rollins}, {Pierbattista}, {Piron}, {Pivato}, {Pletsch},
  {Porter}, {Possenti}, {Rain{\`o}}, {Rando}, {Ransom}, {Ray}, {Razzano},
  {Rea}, {Reimer}, {Reimer}, {Renault}, {Reposeur}, {Ritz}, {Romani}, {Roth},
  {Rousseau}, {Roy}, {Ruan}, {Sartori}, {Saz Parkinson}, {Scargle}, {Schulz},
  {Sgr{\`o}}, {Shannon}, {Siskind}, {Smith}, {Spandre}, {Spinelli}, {Stappers},
  {Strong}, {Suson}, {Takahashi}, {Thayer}, {Thayer}, {Theureau}, {Thompson},
  {Thorsett}, {Tibaldo}, {Tibolla}, {Tinivella}, {Torres}, {Tosti}, {Troja},
  {Uchiyama}, {Usher}, {Vandenbroucke}, {Vasileiou}, {Venter}, {Vianello},
  {Vitale}, {Wang}, {Weltevrede}, {Winer}, {Wolff}, {Wood}, {Wood}, {Wood}, \&
  {Yang}}]{abdo2013}
{Abdo}, A.~A., {Ajello}, M., {Allafort}, A., {et~al.} 2013, \apjs, 208, 17,
  \dodoi{10.1088/0067-0049/208/2/17}

\bibitem[{Akaike(1974)}]{akaike1974}
Akaike, H. 1974, IEEE Transactions on Automatic Control, 19, 716,
  \dodoi{10.1109/TAC.1974.1100705}

\bibitem[{{Anderson} {et~al.}(1994){Anderson}, {Wachter}, {Margon}, {Downes},
  {Blair}, \& {Halpern}}]{anderson1994}
{Anderson}, S.~F., {Wachter}, S., {Margon}, B., {et~al.} 1994, \apj, 436, 319,
  \dodoi{10.1086/174907}

\bibitem[{{Arnaud}(1996)}]{arnaud1996}
{Arnaud}, K.~A. 1996, in Astronomical Society of the Pacific Conference Series,
  Vol. 101, Astronomical Data Analysis Software and Systems V, ed. G.~H.
  {Jacoby} \& J.~{Barnes}, 17

\bibitem[{{Bachetti}(2015{\natexlab{a}})}]{hendrics}
{Bachetti}, M. 2015{\natexlab{a}}, {MaLTPyNT: Quick look timing analysis for
  NuSTAR data}, Astrophysics Source Code Library.
\newblock \doeprint{1502.021}

\bibitem[{{Bachetti}(2015{\natexlab{b}})}]{bachetti2015b}
---. 2015{\natexlab{b}}, {MaLTPyNT: Quick look timing analysis for NuSTAR
  data}, Astrophysics Source Code Library.
\newblock \doeprint{1502.021}

\bibitem[{{Bachetti} \& {Huppenkothen}(2017)}]{bachetti2017}
{Bachetti}, M., \& {Huppenkothen}, D. 2017, ArXiv e-prints.
\newblock \doarXiv{1709.09700}

\bibitem[{Bachetti {et~al.}(2015)Bachetti, Harrison, Cook, Tomsick, Schmid,
  Grefenstette, Barret, Boggs, Christensen, Craig, Fabian, F{\"u}rst, Gandhi,
  Hailey, Kara, Maccarone, Miller, Pottschmidt, Stern, Uttley, Walton, Wilms,
  \& Zhang}]{Bachetti+15}
Bachetti, M., Harrison, F.~A., Cook, R., {et~al.} 2015, ApJ, 800, 109

\bibitem[{{Bahcall} \& {Bahcall}(1972)}]{bahcall1972}
{Bahcall}, J.~N., \& {Bahcall}, N.~A. 1972, \apjl, 178, L1,
  \dodoi{10.1086/181070}

\bibitem[{Barentsen {et~al.}(2019)Barentsen, Hedges, Vinícius, Saunders, gully,
  Hall, Sagear, Barclay, KenMighell, Bell, Zhang, Turtelboom, Berta-Thompson,
  Williams, III, Davies, Vincello, \& Sundaram}]{lightkurve}
Barentsen, G., Hedges, C., Vinícius, Z., {et~al.} 2019, KeplerGO/lightkurve:
  Lightkurve v1.0b30, \dodoi{10.5281/zenodo.2611871}.
\newblock \url{https://doi.org/10.5281/zenodo.2611871}

\bibitem[{{Barret} {et~al.}(2018){Barret}, {Lam Trong}, {den Herder}, {Piro},
  {Cappi}, {Houvelin}, {Kelley}, {Mas-Hesse}, {Mitsuda}, {Paltani}, {Rauw},
  {Rozanska}, {Wilms}, {Bandler}, {Barbera}, {Barcons}, {Bozzo}, {Ceballos},
  {Charles}, {Costantini}, {Decourchelle}, {den Hartog}, {Duband}, {Duval},
  {Fiore}, {Gatti}, {Goldwurm}, {Jackson}, {Jonker}, {Kilbourne}, {Macculi},
  {Mendez}, {Molendi}, {Orleanski}, {Pajot}, {Pointecouteau}, {Porter},
  {Pratt}, {Pr{\^e}le}, {Ravera}, {Sato}, {Schaye}, {Shinozaki}, {Thibert},
  {Valenziano}, {Valette}, {Vink}, {Webb}, {Wise}, {Yamasaki}, {Douchin},
  {Mesnager}, {Pontet}, {Pradines}, {Branduardi-Raymont}, {Bulbul}, {Dadina},
  {Ettori}, {Finoguenov}, {Fukazawa}, {Janiuk}, {Kaastra}, {Mazzotta},
  {Miller}, {Miniutti}, {Naze}, {Nicastro}, {Scioritino}, {Simonescu},
  {Torrejon}, {Frezouls}, {Geoffray}, {Peille}, {Aicardi}, {Andr{\'e}},
  {Daniel}, {Cl{\'e}net}, {Etcheverry}, {Gloaguen}, {Hervet}, {Jolly}, {Ledot},
  {Paillet}, {Schmisser}, {Vella}, {Damery}, {Boyce}, {Dipirro}, {Lotti},
  {Schwander}, {Smith}, {Van Leeuwen}, {van Weers}, {Clerc}, {Cobo}, {Dauser},
  {Kirsch}, {Cucchetti}, {Eckart}, {Ferrando}, \& {Natalucci}}]{athenaXIFU}
{Barret}, D., {Lam Trong}, T., {den Herder}, J.-W., {et~al.} 2018, in Space
  Telescopes and Instrumentation 2018: Ultraviolet to Gamma Ray, Vol. 10699,
  106991G

\bibitem[{{Bellm} {et~al.}(2019){Bellm}, {Kulkarni}, {Graham}, {Dekany},
  {Smith}, {Riddle}, {Masci}, {Helou}, {Prince}, {Adams}, {Barbarino},
  {Barlow}, {Bauer}, {Beck}, {Belicki}, {Biswas}, {Blagorodnova}, {Bodewits},
  {Bolin}, {Brinnel}, {Brooke}, {Bue}, {Bulla}, {Burruss}, {Cenko}, {Chang},
  {Connolly}, {Coughlin}, {Cromer}, {Cunningham}, {De}, {Delacroix}, {Desai},
  {Duev}, {Eadie}, {Farnham}, {Feeney}, {Feindt}, {Flynn}, {Franckowiak},
  {Frederick}, {Fremling}, {Gal-Yam}, {Gezari}, {Giomi}, {Goldstein},
  {Golkhou}, {Goobar}, {Groom}, {Hacopians}, {Hale}, {Henning}, {Ho}, {Hover},
  {Howell}, {Hung}, {Huppenkothen}, {Imel}, {Ip}, {Ivezi{\'c}}, {Jackson},
  {Jones}, {Juric}, {Kasliwal}, {Kaspi}, {Kaye}, {Kelley}, {Kowalski},
  {Kramer}, {Kupfer}, {Landry}, {Laher}, {Lee}, {Lin}, {Lin}, {Lunnan},
  {Giomi}, {Mahabal}, {Mao}, {Miller}, {Monkewitz}, {Murphy}, {Ngeow},
  {Nordin}, {Nugent}, {Ofek}, {Patterson}, {Penprase}, {Porter}, {Rauch},
  {Rebbapragada}, {Reiley}, {Rigault}, {Rodriguez}, {van Roestel}, {Rusholme},
  {van Santen}, {Schulze}, {Shupe}, {Singer}, {Soumagnac}, {Stein}, {Surace},
  {Sollerman}, {Szkody}, {Taddia}, {Terek}, {Van Sistine}, {van Velzen},
  {Vestrand}, {Walters}, {Ward}, {Ye}, {Yu}, {Yan}, \& {Zolkower}}]{bellm2019}
{Bellm}, E.~C., {Kulkarni}, S.~R., {Graham}, M.~J., {et~al.} 2019, \pasp, 131,
  018002, \dodoi{10.1088/1538-3873/aaecbe}

\bibitem[{{Belloni} \& {Hasinger}(1990)}]{belloni1990}
{Belloni}, T., \& {Hasinger}, G. 1990, \aap, 227, L33

\bibitem[{{Belloni} {et~al.}(2005){Belloni}, {Homan}, {Casella}, {van der
  Klis}, {Nespoli}, {Lewin}, {Miller}, \& {M{\'e}ndez}}]{belloni2005}
{Belloni}, T., {Homan}, J., {Casella}, P., {et~al.} 2005, \aap, 440, 207,
  \dodoi{10.1051/0004-6361:20042457}

\bibitem[{{Beloborodov} {et~al.}(2000){Beloborodov}, {Stern}, \&
  {Svensson}}]{beloborodov2000}
{Beloborodov}, A.~M., {Stern}, B.~E., \& {Svensson}, R. 2000, \apj, 535, 158,
  \dodoi{10.1086/308836}

\bibitem[{Bentz(2016)}]{Bentz2016}
Bentz, M.~C. 2016, AGN Reverberation Mapping, ed. H.~M.~J. Boffin, G.~Hussain,
  J.-P. Berger, \& L.~Schmidtobreick (Cham: Springer International Publishing),
  249--266.
\newblock \url{https://doi.org/10.1007/978-3-319-39739-9_13}

\bibitem[{{Beri} {et~al.}(2019){Beri}, {Tetarenko}, {Bahramian}, {Altamirano},
  {Gandhi}, {Sivakoff}, {Degenaar}, {Middleton}, {Wijnands}, {Hern{\'a}ndz
  Santisteban}, \& {Paice}}]{beri2019}
{Beri}, A., {Tetarenko}, B.~E., {Bahramian}, A., {et~al.} 2019, \mnras, 485,
  3064, \dodoi{10.1093/mnras/stz616}

\bibitem[{{Blandford} \& {McKee}(1982)}]{blandford1982}
{Blandford}, R.~D., \& {McKee}, C.~F. 1982, \apj, 255, 419,
  \dodoi{10.1086/159843}

\bibitem[{{Borucki} {et~al.}(2010){Borucki}, {Koch}, {Basri}, {Batalha},
  {Brown}, {Caldwell}, {Caldwell}, {Christensen-Dalsgaard}, {Cochran},
  {DeVore}, {Dunham}, {Dupree}, {Gautier}, {Geary}, {Gilliland}, {Gould},
  {Howell}, {Jenkins}, {Kondo}, {Latham}, {Marcy}, {Meibom}, {Kjeldsen},
  {Lissauer}, {Monet}, {Morrison}, {Sasselov}, {Tarter}, {Boss}, {Brownlee},
  {Owen}, {Buzasi}, {Charbonneau}, {Doyle}, {Fortney}, {Ford}, {Holman},
  {Seager}, {Steffen}, {Welsh}, {Rowe}, {Anderson}, {Buchhave}, {Ciardi},
  {Walkowicz}, {Sherry}, {Horch}, {Isaacson}, {Everett}, {Fischer}, {Torres},
  {Johnson}, {Endl}, {MacQueen}, {Bryson}, {Dotson}, {Haas}, {Kolodziejczak},
  {Van Cleve}, {Chandrasekaran}, {Twicken}, {Quintana}, {Clarke}, {Allen},
  {Li}, {Wu}, {Tenenbaum}, {Verner}, {Bruhweiler}, {Barnes}, \&
  {Prsa}}]{borucki2010}
{Borucki}, W.~J., {Koch}, D., {Basri}, G., {et~al.} 2010, Science, 327, 977,
  \dodoi{10.1126/science.1185402}

\bibitem[{{Bradt} {et~al.}(1993){Bradt}, {Rothschild}, \&
  {Swank}}]{Bradtetal93}
{Bradt}, H.~V., {Rothschild}, R.~E., \& {Swank}, J.~H. 1993, \aaps, 97, 355

\bibitem[{{Brown} {et~al.}(2011){Brown}, {Latham}, {Everett}, \&
  {Esquerdo}}]{brown2011}
{Brown}, T.~M., {Latham}, D.~W., {Everett}, M.~E., \& {Esquerdo}, G.~A. 2011,
  \aj, 142, 112, \dodoi{10.1088/0004-6256/142/4/112}

\bibitem[{{Brumback} {et~al.}(2018){Brumback}, {Hickox}, {Bachetti},
  {Ballhausen}, {F{\"u}rst}, {Pike}, {Pottschmidt}, {Tomsick}, \&
  {Wilms}}]{brumback2018}
{Brumback}, M.~C., {Hickox}, R.~C., {Bachetti}, M., {et~al.} 2018, \apj, 861,
  L7, \dodoi{10.3847/2041-8213/aacd13}

\bibitem[{{Buccheri} {et~al.}(1983){Buccheri}, {Bennett}, {Bignami}, {Bloemen},
  {Boriakoff}, {Caraveo}, {Hermsen}, {Kanbach}, {Manchester}, {Masnou},
  {Mayer-Hasselwander}, {{\"O}zel}, {Paul}, {Sacco}, {Scarsi}, \&
  {Strong}}]{buccheri1983}
{Buccheri}, R., {Bennett}, K., {Bignami}, G.~F., {et~al.} 1983, \aap, 128, 245

\bibitem[{Burke {et~al.}(2018)Burke, Laurino, dtnguyen2, Budynkiewicz,
  Aldcroft, Siemiginowska, wmclaugh, Sipocz, Deil, \& Leinweber}]{sherpa}
Burke, D., Laurino, O., dtnguyen2, {et~al.} 2018, sherpa/sherpa: Sherpa 4.10.1,
  \dodoi{10.5281/zenodo.1463962}.
\newblock \url{https://doi.org/10.5281/zenodo.1463962}

\bibitem[{Carpenter {et~al.}(2017)Carpenter, Gelman, Hoffman, Lee, Goodrich,
  Betancourt, Brubaker, Guo, Li, \& Riddell}]{stan}
Carpenter, B., Gelman, A., Hoffman, M., {et~al.} 2017, Journal of Statistical
  Software, Articles, 76, 1, \dodoi{10.18637/jss.v076.i01}

\bibitem[{{Cash}(1979)}]{cash1979}
{Cash}, W. 1979, \apj, 228, 939, \dodoi{10.1086/156922}

\bibitem[{{Charbonneau} {et~al.}(2000){Charbonneau}, {Brown}, {Latham}, \&
  {Mayor}}]{charbonneau2000}
{Charbonneau}, D., {Brown}, T.~M., {Latham}, D.~W., \& {Mayor}, M. 2000, \apjl,
  529, L45, \dodoi{10.1086/312457}

\bibitem[{{Cheng} {et~al.}(1995){Cheng}, {Vrtilek}, \& {Raymond}}]{cheng1995}
{Cheng}, F.~H., {Vrtilek}, S.~D., \& {Raymond}, J.~C. 1995, \apj, 452, 825,
  \dodoi{10.1086/176351}

\bibitem[{{Coughlin} {et~al.}(2016){Coughlin}, {Mullally}, {Thompson}, {Rowe},
  {Burke}, {Latham}, {Batalha}, {Ofir}, {Quarles}, {Henze}, {Wolfgang},
  {Caldwell}, {Bryson}, {Shporer}, {Catanzarite}, {Akeson}, {Barclay},
  {Borucki}, {Boyajian}, {Campbell}, {Christiansen}, {Girouard}, {Haas},
  {Howell}, {Huber}, {Jenkins}, {Li}, {Patil-Sabale}, {Quintana}, {Ramirez},
  {Seader}, {Smith}, {Tenenbaum}, {Twicken}, \& {Zamudio}}]{coughlin2016}
{Coughlin}, J.~L., {Mullally}, F., {Thompson}, S.~E., {et~al.} 2016, \apjs,
  224, 12, \dodoi{10.3847/0067-0049/224/1/12}

\bibitem[{{Davidsen} {et~al.}(1972){Davidsen}, {Henry}, {Middleditch}, \&
  {Smith}}]{davidsen1972}
{Davidsen}, A., {Henry}, J.~P., {Middleditch}, J., \& {Smith}, H.~E. 1972,
  \apjl, 177, L97, \dodoi{10.1086/181060}

\bibitem[{{Di Mauro}(2016)}]{dimauro2016}
{Di Mauro}, M.~P. 2016, in Frontier Research in Astrophysics II, 29

\bibitem[{Feigelson {et~al.}(2018)Feigelson, Babu, \& Caceres}]{feigelson2018}
Feigelson, E.~D., Babu, G.~J., \& Caceres, G.~A. 2018, Frontiers in Physics, 6,
  80, \dodoi{10.3389/fphy.2018.00080}

\bibitem[{Forbrich {et~al.}(2017)Forbrich, Reid, Menten, Rivilla, Wolk, Rau, \&
  Chandler}]{Forbrich_2017}
Forbrich, J., Reid, M.~J., Menten, K.~M., {et~al.} 2017, The Astrophysical
  Journal, 844, 109, \dodoi{10.3847/1538-4357/aa7aa4}

\bibitem[{Foreman-Mackey(2016)}]{corner}
Foreman-Mackey, D. 2016, The Journal of Open Source Software, 24,
  \dodoi{10.21105/joss.00024}

\bibitem[{{Foreman-Mackey} {et~al.}(2017){Foreman-Mackey}, {Agol},
  {Ambikasaran}, \& {Angus}}]{foremanmackey2017}
{Foreman-Mackey}, D., {Agol}, E., {Ambikasaran}, S., \& {Angus}, R. 2017, \aj,
  154, 220, \dodoi{10.3847/1538-3881/aa9332}

\bibitem[{{Foreman-Mackey} {et~al.}(2013){Foreman-Mackey}, {Hogg}, {Lang}, \&
  {Goodman}}]{emcee}
{Foreman-Mackey}, D., {Hogg}, D.~W., {Lang}, D., \& {Goodman}, J. 2013, \pasp,
  125, 306, \dodoi{10.1086/670067}

\bibitem[{{Forman} {et~al.}(1972){Forman}, {Jones}, \& {Liller}}]{forman1972}
{Forman}, W., {Jones}, C.~A., \& {Liller}, W. 1972, \apjl, 177, L103,
  \dodoi{10.1086/181061}

\bibitem[{F{\"u}rst {et~al.}(2013)F{\"u}rst, Grefenstette, Staubert, Tomsick,
  Bachetti, Barret, Bellm, Boggs, Chenevez, Christensen, Craig, Hailey,
  Harrison, Klochkov, Madsen, Pottschmidt, Stern, Walton, Wilms, \&
  Zhang}]{Fuerst13}
F{\"u}rst, F., Grefenstette, B.~W., Staubert, R., {et~al.} 2013, The
  Astrophysical Journal, 779, 69

\bibitem[{Gelman \& Rubin(1992)}]{gelman1992}
Gelman, A., \& Rubin, D.~B. 1992, Statist. Sci., 7, 457,
  \dodoi{10.1214/ss/1177011136}

\bibitem[{{Gendreau} {et~al.}(2016){Gendreau}, {Arzoumanian}, {Adkins},
  {Albert}, {Anders}, {Aylward}, {Baker}, {Balsamo}, {Bamford}, {Benegalrao},
  {Berry}, {Bhalwani}, {Black}, {Blaurock}, {Bronke}, {Brown}, {Budinoff},
  {Cantwell}, {Cazeau}, {Chen}, {Clement}, {Colangelo}, {Coleman},
  {Coopersmith}, {Dehaven}, {Doty}, {Egan}, {Enoto}, {Fan}, {Ferro}, {Foster},
  {Galassi}, {Gallo}, {Green}, {Grosh}, {Ha}, {Hasouneh}, {Heefner}, {Hestnes},
  {Hoge}, {Jacobs}, {J{\o}rgensen}, {Kaiser}, {Kellogg}, {Kenyon}, {Koenecke},
  {Kozon}, {LaMarr}, {Lambertson}, {Larson}, {Lentine}, {Lewis}, {Lilly},
  {Liu}, {Malonis}, {Manthripragada}, {Markwardt}, {Matonak}, {Mcginnis},
  {Miller}, {Mitchell}, {Mitchell}, {Mohammed}, {Monroe}, {Montt de Garcia},
  {Mul{\'e}}, {Nagao}, {Ngo}, {Norris}, {Norwood}, {Novotka}, {Okajima},
  {Olsen}, {Onyeachu}, {Orosco}, {Peterson}, {Pevear}, {Pham}, {Pollard},
  {Pope}, {Powers}, {Powers}, {Price}, {Prigozhin}, {Ramirez}, {Reid},
  {Remillard}, {Rogstad}, {Rosecrans}, {Rowe}, {Sager}, {Sanders}, {Savadkin},
  {Saylor}, {Schaeffer}, {Schweiss}, {Semper}, {Serlemitsos}, {Shackelford},
  {Soong}, {Struebel}, {Vezie}, {Villasenor}, {Winternitz}, {Wofford},
  {Wright}, {Yang}, \& {Yu}}]{gendreau2016}
{Gendreau}, K.~C., {Arzoumanian}, Z., {Adkins}, P.~W., {et~al.} 2016, in
  \procspie, Vol. 9905, Space Telescopes and Instrumentation 2016: Ultraviolet
  to Gamma Ray, 99051H

\bibitem[{{Giacconi} {et~al.}(1973){Giacconi}, {Gursky}, {Kellogg}, {Levinson},
  {Schreier}, \& {Tananbaum}}]{giacconi1973}
{Giacconi}, R., {Gursky}, H., {Kellogg}, E., {et~al.} 1973, \apj, 184, 227,
  \dodoi{10.1086/152321}

\bibitem[{Girod {et~al.}(2001)Girod, Rabenstein, \& Stenger}]{girod2001signals}
Girod, B., Rabenstein, R., \& Stenger, A. 2001, Signals and systems (Wiley).
\newblock \url{https://books.google.com/books?id=DoseAQAAIAAJ}

\bibitem[{{Graham} {et~al.}(2019){Graham}, {Kulkarni}, {Bellm}, {Adams},
  {Barbarino}, {Blagorodnova}, {Bodewits}, {Bolin}, {Brady}, {Cenko}, {Chang},
  {Coughlin}, {De}, {Eadie}, {Farnham}, {Feindt}, {Franckowiak}, {Fremling},
  {Gal-yam}, {Gezari}, {Ghosh}, {Goldstein}, {Golkhou}, {Goobar}, {Ho},
  {Huppenkothen}, {Ivezic}, {Jones}, {Juric}, {Kaplan}, {Kasliwal}, {Kelley},
  {Kupfer}, {Lee}, {Lin}, {Lunnan}, {Mahabal}, {Miller}, {Ngeow}, {Nugent},
  {Ofek}, {Prince}, {Rauch}, {van Roestel}, {Schulze}, {Singer}, {Sollerman},
  {Taddia}, {Yan}, {Ye}, {Yu}, {Andreoni}, {Barlow}, {Bauer}, {Beck},
  {Belicki}, {Biswas}, {Brinnel}, {Brooke}, {Bue}, {Bulla}, {Burdge},
  {Burruss}, {Connolly}, {Cromer}, {Cunningham}, {Dekany}, {Delacroix},
  {Desai}, {Duev}, {Hacopians}, {Hale}, {Helou}, {Henning}, {Hover},
  {Hillenbrand}, {Howell}, {Hung}, {Imel}, {Ip}, {Jackson}, {Kaspi}, {Kaye},
  {Kowalski}, {Kramer}, {Kuhn}, {Landry}, {Laher}, {Mao}, {Masci}, {Monkewitz},
  {Murphy}, {Nordin}, {Patterson}, {Penprase}, {Porter}, {Rebbapragada},
  {Reiley}, {Riddle}, {Rigault}, {Rodriguez}, {Rusholme}, {van Santen},
  {Shupe}, {Smith}, {Soumagnac}, {Stein}, {Surace}, {Szkody}, {Terek}, {van
  Sistine}, {van Velzen}, {Vestrand}, {Walters}, {Ward}, {Zhang}, \&
  {Zolkower}}]{graham2019}
{Graham}, M.~J., {Kulkarni}, S.~R., {Bellm}, E.~C., {et~al.} 2019, arXiv
  e-prints.
\newblock \doarXiv{1902.01945}

\bibitem[{{Groth}(1975)}]{Groth1975}
{Groth}, E.~J. 1975, \apjs, 29, 285, \dodoi{10.1086/190343}

\bibitem[{{Guidorzi} {et~al.}(2016){Guidorzi}, {Dichiara}, \&
  {Amati}}]{guidorzi2016}
{Guidorzi}, C., {Dichiara}, S., \& {Amati}, L. 2016, \aap, 589, A98,
  \dodoi{10.1051/0004-6361/201527642}

\bibitem[{Harrison {et~al.}(2013)Harrison, Craig, Christensen, Hailey, Zhang,
  Boggs, Stern, Cook, Forster, Giommi, Grefenstette, Kim, Kitaguchi, Koglin,
  Madsen, Mao, Miyasaka, Mori, Perri, Pivovaroff, Puccetti, Rana, Westergaard,
  Willis, Zoglauer, An, Bachetti, Barri{\`e}re, Bellm, Bhalerao, Brejnholt,
  Fuerst, Liebe, Markwardt, Nynka, Vogel, Walton, Wik, Alexander, Cominsky,
  Hornschemeier, Hornstrup, Kaspi, Madejski, Matt, Molendi, Smith, Tomsick,
  Ajello, Ballantyne, Balokovi{\'c}, Barret, Bauer, Blandford, Brandt,
  Brenneman, Chiang, Chakrabarty, Chenevez, Comastri, Dufour, Elvis, Fabian,
  Farrah, Fryer, Gotthelf, Grindlay, Helfand, Krivonos, Meier, Miller,
  Natalucci, Ogle, Ofek, Ptak, Reynolds, Rigby, Tagliaferri, Thorsett,
  Treister, \& Urry}]{nustar13}
Harrison, F.~A., Craig, W.~W., Christensen, F.~E., {et~al.} 2013, ApJ, 770, 103

\bibitem[{{Heida} {et~al.}(2017){Heida}, {Jonker}, {Torres}, \&
  {Chiavassa}}]{Heidaetal17}
{Heida}, M., {Jonker}, P.~G., {Torres}, M.~A.~P., \& {Chiavassa}, A. 2017,
  \apj, 846, 132, \dodoi{10.3847/1538-4357/aa85df}

\bibitem[{{Henry} {et~al.}(2000){Henry}, {Marcy}, {Butler}, \&
  {Vogt}}]{henry2000}
{Henry}, G.~W., {Marcy}, G.~W., {Butler}, R.~P., \& {Vogt}, S.~S. 2000, \apjl,
  529, L41, \dodoi{10.1086/312458}

\bibitem[{{Hewish} {et~al.}(1968){Hewish}, {Bell}, {Pilkington}, {Scott}, \&
  {Collins}}]{hewish1968}
{Hewish}, A., {Bell}, S.~J., {Pilkington}, J.~D.~H., {Scott}, P.~F., \&
  {Collins}, R.~A. 1968, \nat, 217, 709, \dodoi{10.1038/217709a0}

\bibitem[{{Houck} \& {Denicola}(2000)}]{houck2000}
{Houck}, J.~C., \& {Denicola}, L.~A. 2000, in Astronomical Society of the
  Pacific Conference Series, Vol. 216, Astronomical Data Analysis Software and
  Systems IX, ed. N.~{Manset}, C.~{Veillet}, \& D.~{Crabtree}, 591

\bibitem[{Hunter(2007)}]{matplotlib}
Hunter, J.~D. 2007, Computing in Science Engineering, 9, 90,
  \dodoi{10.1109/MCSE.2007.55}

\bibitem[{{Huppenkothen} \& {Bachetti}(2017)}]{huppenkothen2017}
{Huppenkothen}, D., \& {Bachetti}, M. 2017, ArXiv e-prints.
\newblock \doarXiv{1709.09666}

\bibitem[{{Huppenkothen} {et~al.}(2013){Huppenkothen}, {Watts}, {Uttley}, {van
  der Horst}, {van der Klis}, {Kouveliotou}, {G{\"o}{\v g}{\"u}{\c s}},
  {Granot}, {Vaughan}, \& {Finger}}]{huppenkothen2013}
{Huppenkothen}, D., {Watts}, A.~L., {Uttley}, P., {et~al.} 2013, \apj, 768, 87,
  \dodoi{10.1088/0004-637X/768/1/87}

\bibitem[{{Hynes} {et~al.}(2003){Hynes}, {Steeghs}, {Casares}, {Charles}, \&
  {O'Brien}}]{Hynesetal03}
{Hynes}, R.~I., {Steeghs}, D., {Casares}, J., {Charles}, P.~A., \& {O'Brien},
  K. 2003, \apjl, 583, L95, \dodoi{10.1086/368108}

\bibitem[{{Igna} \& {Leahy}(2011)}]{igna2011}
{Igna}, C.~D., \& {Leahy}, D.~A. 2011, \mnras, 418, 2283,
  \dodoi{10.1111/j.1365-2966.2011.19550.x}

\bibitem[{{Inglis} {et~al.}(2016){Inglis}, {Ireland}, {Dennis}, {Hayes}, \&
  {Gallagher}}]{inglis2016}
{Inglis}, A.~R., {Ireland}, J., {Dennis}, B.~R., {Hayes}, L., \& {Gallagher},
  P. 2016, \apj, 833, 284, \dodoi{10.3847/1538-4357/833/2/284}

\bibitem[{{Ingram} \& {van der Klis}(2015)}]{ingram2015}
{Ingram}, A., \& {van der Klis}, M. 2015, \mnras, 446, 3516,
  \dodoi{10.1093/mnras/stu2373}

\bibitem[{{Israel} {et~al.}(2005){Israel}, {Belloni}, {Stella}, {Rephaeli},
  {Gruber}, {Casella}, {Dall'Osso}, {Rea}, {Persic}, \&
  {Rothschild}}]{israel2005}
{Israel}, G.~L., {Belloni}, T., {Stella}, L., {et~al.} 2005, \apjl, 628, L53,
  \dodoi{10.1086/432615}

\bibitem[{{Ivezi{\'c}} {et~al.}(2019){Ivezi{\'c}}, {Kahn}, {Tyson}, {Abel},
  {Acosta}, {Allsman}, {Alonso}, {AlSayyad}, {Anderson}, {Andrew}, {Angel},
  {Angeli}, {Ansari}, {Antilogus}, {Araujo}, {Armstrong}, {Arndt}, {Astier},
  {Aubourg}, {Auza}, {Axelrod}, {Bard}, {Barr}, {Barrau}, {Bartlett}, {Bauer},
  {Bauman}, {Baumont}, {Bechtol}, {Bechtol}, {Becker}, {Becla}, {Beldica},
  {Bellavia}, {Bianco}, {Biswas}, {Blanc}, {Blazek}, {Bland ford}, {Bloom},
  {Bogart}, {Bond}, {Booth}, {Borgland}, {Borne}, {Bosch}, {Boutigny},
  {Brackett}, {Bradshaw}, {Nielsen Brand t}, {Brown}, {Bullock}, {Burchat},
  {Burke}, {Cagnoli}, {Calabrese}, {Callahan}, {Callen}, {Carlin}, {Carlson},
  {Chand rasekharan}, {Charles-Emerson}, {Chesley}, {Cheu}, {Chiang}, {Chiang},
  {Chirino}, {Chow}, {Ciardi}, {Claver}, {Cohen-Tanugi}, {Cockrum}, {Coles},
  {Connolly}, {Cook}, {Cooray}, {Covey}, {Cribbs}, {Cui}, {Cutri}, {Daly},
  {Daniel}, {Daruich}, {Daubard}, {Daues}, {Dawson}, {Delgado}, {Dellapenna},
  {de Peyster}, {de Val-Borro}, {Digel}, {Doherty}, {Dubois},
  {Dubois-Felsmann}, {Durech}, {Economou}, {Eifler}, {Eracleous}, {Emmons},
  {Fausti Neto}, {Ferguson}, {Figueroa}, {Fisher-Levine}, {Focke}, {Foss},
  {Frank}, {Freemon}, {Gangler}, {Gawiser}, {Geary}, {Gee}, {Geha}, {Gessner},
  {Gibson}, {Gilmore}, {Glanzman}, {Glick}, {Goldina}, {Goldstein}, {Goodenow},
  {Graham}, {Gressler}, {Gris}, {Guy}, {Guyonnet}, {Haller}, {Harris},
  {Hascall}, {Haupt}, {Hernand ez}, {Herrmann}, {Hileman}, {Hoblitt},
  {Hodgson}, {Hogan}, {Howard}, {Huang}, {Huffer}, {Ingraham}, {Innes},
  {Jacoby}, {Jain}, {Jammes}, {Jee}, {Jenness}, {Jernigan}, {Jevremovi{\'c}},
  {Johns}, {Johnson}, {Johnson}, {Jones}, {Juramy-Gilles}, {Juri{\'c}},
  {Kalirai}, {Kallivayalil}, {Kalmbach}, {Kantor}, {Karst}, {Kasliwal},
  {Kelly}, {Kessler}, {Kinnison}, {Kirkby}, {Knox}, {Kotov}, {Krabbendam},
  {Krughoff}, {Kub{\'a}nek}, {Kuczewski}, {Kulkarni}, {Ku}, {Kurita}, {Lage},
  {Lambert}, {Lange}, {Langton}, {Le Guillou}, {Levine}, {Liang}, {Lim},
  {Lintott}, {Long}, {Lopez}, {Lotz}, {Lupton}, {Lust}, {MacArthur}, {Mahabal},
  {Mand elbaum}, {Markiewicz}, {Marsh}, {Marshall}, {Marshall}, {May},
  {McKercher}, {McQueen}, {Meyers}, {Migliore}, {Miller}, {Mills}, {Miraval},
  {Moeyens}, {Moolekamp}, {Monet}, {Moniez}, {Monkewitz}, {Montgomery},
  {Morrison}, {Mueller}, {Muller}, {Mu{\~n}oz Arancibia}, {Neill}, {Newbry},
  {Nief}, {Nomerotski}, {Nordby}, {O{\textquoteright}Connor}, {Oliver},
  {Olivier}, {Olsen}, {O{\textquoteright}Mullane}, {Ortiz}, {Osier}, {Owen},
  {Pain}, {Palecek}, {Parejko}, {Parsons}, {Pease}, {Peterson}, {Peterson},
  {Petravick}, {Libby Petrick}, {Petry}, {Pierfederici}, {Pietrowicz}, {Pike},
  {Pinto}, {Plante}, {Plate}, {Plutchak}, {Price}, {Prouza}, {Radeka},
  {Rajagopal}, {Rasmussen}, {Regnault}, {Reil}, {Reiss}, {Reuter}, {Ridgway},
  {Riot}, {Ritz}, {Robinson}, {Roby}, {Roodman}, {Rosing}, {Roucelle},
  {Rumore}, {Russo}, {Saha}, {Sassolas}, {Schalk}, {Schellart}, {Schindler},
  {Schmidt}, {Schneider}, {Schneider}, {Schoening}, {Schumacher}, {Schwamb},
  {Sebag}, {Selvy}, {Sembroski}, {Seppala}, {Serio}, {Serrano}, {Shaw},
  {Shipsey}, {Sick}, {Silvestri}, {Slater}, {Smith}, {Smith}, {Sobhani},
  {Soldahl}, {Storrie-Lombardi}, {Stover}, {Strauss}, {Street}, {Stubbs},
  {Sullivan}, {Sweeney}, {Swinbank}, {Szalay}, {Takacs}, {Tether}, {Thaler},
  {Thayer}, {Thomas}, {Thornton}, {Thukral}, {Tice}, {Trilling}, {Turri}, {Van
  Berg}, {Vanden Berk}, {Vetter}, {Virieux}, {Vucina}, {Wahl}, {Walkowicz},
  {Walsh}, {Walter}, {Wang}, {Wang}, {Warner}, {Wiecha}, {Willman}, {Winters},
  {Wittman}, {Wolff}, {Wood-Vasey}, {Wu}, {Xin}, {Yoachim}, \&
  {Zhan}}]{lsst2019}
{Ivezi{\'c}}, {\v{Z}}., {Kahn}, S.~M., {Tyson}, J.~A., {et~al.} 2019, \apj,
  873, 111, \dodoi{10.3847/1538-4357/ab042c}

\bibitem[{{Jahoda} {et~al.}(1996){Jahoda}, {Swank}, {Giles}, {Stark},
  {Strohmayer}, {Zhang}, \& {Morgan}}]{Jahodaetal96}
{Jahoda}, K., {Swank}, J.~H., {Giles}, A.~B., {et~al.} 1996, in \procspie, Vol.
  2808, EUV, X-Ray, and Gamma-Ray Instrumentation for Astronomy VII, ed. O.~H.
  {Siegmund} \& M.~A. {Gummin}, 59--70

\bibitem[{Jaisawal {et~al.}(2018)Jaisawal, Naik, Gupta, Chenevez, \&
  Epili}]{jaisawal2018}
Jaisawal, G.~K., Naik, S., Gupta, S., Chenevez, J., \& Epili, P. 2018, Monthly
  Notices of the Royal Astronomical Society, 478, 448,
  \dodoi{10.1093/mnras/sty1049}

\bibitem[{{Jansen} {et~al.}(2001){Jansen}, {Lumb}, {Altieri}, {Clavel}, {Ehle},
  {Erd}, {Gabriel}, {Guainazzi}, {Gondoin}, {Much}, {Munoz}, {Santos},
  {Schartel}, {Texier}, \& {Vacanti}}]{jansen2001}
{Jansen}, F., {Lumb}, D., {Altieri}, B., {et~al.} 2001, \aap, 365, L1,
  \dodoi{10.1051/0004-6361:20000036}

\bibitem[{Jones {et~al.}(2001--)Jones, Oliphant, Peterson, {et~al.}}]{scipy}
Jones, E., Oliphant, T., Peterson, P., {et~al.} 2001--, {SciPy}: Open source
  scientific tools for {Python}.
\newblock \url{http://www.scipy.org/}

\bibitem[{{Kelly} {et~al.}(2014){Kelly}, {Becker}, {Sobolewska},
  {Siemiginowska}, \& {Uttley}}]{kelly2014}
{Kelly}, B.~C., {Becker}, A.~C., {Sobolewska}, M., {Siemiginowska}, A., \&
  {Uttley}, P. 2014, \apj, 788, 33, \dodoi{10.1088/0004-637X/788/1/33}

\bibitem[{{Kennedy} {et~al.}(2018){Kennedy}, {Clark}, {Voisin}, \&
  {Breton}}]{kennedy2018}
{Kennedy}, M.~R., {Clark}, C.~J., {Voisin}, G., \& {Breton}, R.~P. 2018,
  \mnras, 477, 1120, \dodoi{10.1093/mnras/sty731}

\bibitem[{Lam {et~al.}(2015)Lam, Pitrou, \& Seibert}]{numba}
Lam, S.~K., Pitrou, A., \& Seibert, S. 2015, in Proceedings of the Second
  Workshop on the LLVM Compiler Infrastructure in HPC, LLVM '15 (New York, NY,
  USA: ACM), 7:1--7:6.
\newblock \url{http://doi.acm.org/10.1145/2833157.2833162}

\bibitem[{{Leahy}(1987)}]{leahy1987}
{Leahy}, D.~A. 1987, \aap, 180, 275

\bibitem[{{Leahy} \& {Abdallah}(2014)}]{leahy2014}
{Leahy}, D.~A., \& {Abdallah}, M.~H. 2014, \apj, 793, 79,
  \dodoi{10.1088/0004-637X/793/2/79}

\bibitem[{{Leahy} {et~al.}(1983{\natexlab{a}}){Leahy}, {Darbro}, {Elsner},
  {Weisskopf}, {Kahn}, {Sutherland}, \& {Grindlay}}]{leahy1983}
{Leahy}, D.~A., {Darbro}, W., {Elsner}, R.~F., {et~al.} 1983{\natexlab{a}},
  \apj, 266, 160, \dodoi{10.1086/160766}

\bibitem[{{Leahy} {et~al.}(1983{\natexlab{b}}){Leahy}, {Elsner}, \&
  {Weisskopf}}]{leahy1983b}
{Leahy}, D.~A., {Elsner}, R.~F., \& {Weisskopf}, M.~C. 1983{\natexlab{b}},
  \apj, 272, 256, \dodoi{10.1086/161288}

\bibitem[{Lorimer(2008)}]{Lorimer2008}
Lorimer, D.~R. 2008, Living Reviews in Relativity, 11, 8,
  \dodoi{10.12942/lrr-2008-8}

\bibitem[{{Ludlam} {et~al.}(2015){Ludlam}, {Miller}, \&
  {Cackett}}]{Ludlametal15}
{Ludlam}, R.~M., {Miller}, J.~M., \& {Cackett}, E.~M. 2015, \apj, 806, 262,
  \dodoi{10.1088/0004-637X/806/2/262}

\bibitem[{{Maccarone}(2013)}]{maccarone2013}
{Maccarone}, T.~J. 2013, \mnras, 435, 3547, \dodoi{10.1093/mnras/stt1546}

\bibitem[{{Maccarone} \& {Coppi}(2002)}]{maccarone2002}
{Maccarone}, T.~J., \& {Coppi}, P.~S. 2002, \mnras, 336, 817,
  \dodoi{10.1046/j.1365-8711.2002.05807.x}

\bibitem[{{Maccarone} \& {Schnittman}(2005)}]{maccarone2005}
{Maccarone}, T.~J., \& {Schnittman}, J.~D. 2005, \mnras, 357, 12,
  \dodoi{10.1111/j.1365-2966.2004.08615.x}

\bibitem[{{Miyamoto} {et~al.}(1992){Miyamoto}, {Kitamoto}, {Iga}, {Negoro}, \&
  {Terada}}]{miyamoto1992}
{Miyamoto}, S., {Kitamoto}, S., {Iga}, S., {Negoro}, H., \& {Terada}, K. 1992,
  \apjl, 391, L21, \dodoi{10.1086/186389}

\bibitem[{{Mu{\~n}oz-Darias} {et~al.}(2008){Mu{\~n}oz-Darias}, {Casares}, \&
  {Mart{\'{\i}}nez-Pais}}]{MunozDariasetal08}
{Mu{\~n}oz-Darias}, T., {Casares}, J., \& {Mart{\'{\i}}nez-Pais}, I.~G. 2008,
  \mnras, 385, 2205, \dodoi{10.1111/j.1365-2966.2008.12987.x}

\bibitem[{{Nowak} {et~al.}(1999){Nowak}, {Dove}, {Vaughan}, {Wilms}, \&
  {Begelman}}]{nowak1999}
{Nowak}, M.~A., {Dove}, J.~B., {Vaughan}, B.~A., {Wilms}, J., \& {Begelman},
  M.~C. 1999, Nuclear Physics B Proceedings Supplements, 69, 302,
  \dodoi{10.1016/S0920-5632(98)00229-1}

\bibitem[{{Papitto} {et~al.}(2019){Papitto}, {Ambrosino}, {Stella}, {Torres},
  {Coti Zelati}, {Ghedina}, {Meddi}, {Sanna}, {Casella}, {Dallilar},
  {Eikenberry}, {Israel}, {Onori}, {Piranomonte}, {Bozzo}, {Burderi},
  {Campana}, {de Martino}, {Di Salvo}, {Ferrigno}, {Rea}, {Riggio}, \&
  {Zampieri}}]{papitto2019}
{Papitto}, A., {Ambrosino}, F., {Stella}, L., {et~al.} 2019, arXiv e-prints,
  arXiv:1904.10433.
\newblock \doarXiv{1904.10433}

\bibitem[{{Pike} {et~al.}(2019){Pike}, {Harrison}, {Bachetti}, {Brumback},
  {F{\"u}rst}, {Madsen}, {Pottschmidt}, {Tomsick}, \& {Wilms}}]{pike2019}
{Pike}, S.~N., {Harrison}, F.~A., {Bachetti}, M., {et~al.} 2019, \apj, 875,
  144, \dodoi{10.3847/1538-4357/ab0f2b}

\bibitem[{{P}roject {J}upyter {et~al.}(2018){P}roject {J}upyter, {M}atthias
  {B}ussonnier, {J}essica {F}orde, {J}eremy {F}reeman, {B}rian {G}ranger, {T}im
  {H}ead, {C}hris {H}oldgraf, {K}yle {K}elley, {G}ladys {N}alvarte, {A}ndrew
  {O}sheroff, {P}acer, {Y}uvi {P}anda, {F}ernando {P}erez, {B}enjamin~{R}agan
  {K}elley, \& {C}arol {W}illing}]{project_jupyter-proc-scipy-2018}
{P}roject {J}upyter, {M}atthias {B}ussonnier, {J}essica {F}orde, {et~al.} 2018,
  in {P}roceedings of the 17th {P}ython in {S}cience {C}onference, ed. {F}atih
  {A}kici, {D}avid {L}ippa, {D}illon {N}iederhut, \& M.~{P}acer, 113 -- 120

\bibitem[{{Protassov} {et~al.}(2002){Protassov}, {van Dyk}, {Connors},
  {Kashyap}, \& {Siemiginowska}}]{protassov2002}
{Protassov}, R., {van Dyk}, D.~A., {Connors}, A., {Kashyap}, V.~L., \&
  {Siemiginowska}, A. 2002, \apj, 571, 545, \dodoi{10.1086/339856}

\bibitem[{{Ray} {et~al.}(2018){Ray}, {Arzoumanian}, {Brandt}, {Burns},
  {Chakrabarty}, {Feroci}, {Gendreau}, {Gevin}, {Hernanz}, {Jenke}, {Kenyon},
  {G{\'a}lvez}, {Maccarone}, {Okajima}, {Remillard}, {Schanne}, {Tenzer},
  {Vacchi}, {Wilson-Hodge}, {Winter}, {Zane}, {Ballantyne}, {Bozzo},
  {Brenneman}, {Cackett}, {De Rosa}, {Goldstein}, {Hartmann}, {McDonald},
  {Stevens}, {Tomsick}, {Watts}, {Wood}, \& {Zoghbi}}]{strobex18}
{Ray}, P.~S., {Arzoumanian}, Z., {Brandt}, S., {et~al.} 2018, in Society of
  Photo-Optical Instrumentation Engineers (SPIE) Conference Series, Vol. 10699,
  1069919

\bibitem[{{Revnivtsev} {et~al.}(1999){Revnivtsev}, {Gilfanov}, \&
  {Churazov}}]{Revnivtsevetal99}
{Revnivtsev}, M., {Gilfanov}, M., \& {Churazov}, E. 1999, \aap, 347, L23.
\newblock \doarXiv{astro-ph/9906198}

\bibitem[{{Reynolds} {et~al.}(1997){Reynolds}, {Quaintrell}, {Still}, {Roche},
  {Chakrabarty}, \& {Levine}}]{reynolds1997}
{Reynolds}, A.~P., {Quaintrell}, H., {Still}, M.~D., {et~al.} 1997, \mnras,
  288, 43, \dodoi{10.1093/mnras/288.1.43}

\bibitem[{{Ricker} {et~al.}(2015){Ricker}, {Winn}, {Vanderspek}, {Latham},
  {Bakos}, {Bean}, {Berta-Thompson}, {Brown}, {Buchhave}, {Butler}, {Butler},
  {Chaplin}, {Charbonneau}, {Christensen-Dalsgaard}, {Clampin}, {Deming},
  {Doty}, {De Lee}, {Dressing}, {Dunham}, {Endl}, {Fressin}, {Ge}, {Henning},
  {Holman}, {Howard}, {Ida}, {Jenkins}, {Jernigan}, {Johnson}, {Kaltenegger},
  {Kawai}, {Kjeldsen}, {Laughlin}, {Levine}, {Lin}, {Lissauer}, {MacQueen},
  {Marcy}, {McCullough}, {Morton}, {Narita}, {Paegert}, {Palle}, {Pepe},
  {Pepper}, {Quirrenbach}, {Rinehart}, {Sasselov}, {Sato}, {Seager},
  {Sozzetti}, {Stassun}, {Sullivan}, {Szentgyorgyi}, {Torres}, {Udry}, \&
  {Villasenor}}]{ricker2015}
{Ricker}, G.~R., {Winn}, J.~N., {Vanderspek}, R., {et~al.} 2015, Journal of
  Astronomical Telescopes, Instruments, and Systems, 1, 014003,
  \dodoi{10.1117/1.JATIS.1.1.014003}

\bibitem[{Scaringi {et~al.}(2017)Scaringi, Maccarone, D'Angelo, Knigge, \&
  Groot}]{scaringi2017}
Scaringi, S., Maccarone, T.~J., D'Angelo, C., Knigge, C., \& Groot, P.~J. 2017,
  Nature, 552, 210 EP

\bibitem[{Scaringi {et~al.}(2015)Scaringi, Maccarone, K{\"o}rding, Knigge,
  Vaughan, Marsh, Aranzana, Dhillon, \& Barros}]{scaringi2015}
Scaringi, S., Maccarone, T.~J., K{\"o}rding, E., {et~al.} 2015, Science
  Advances, 1, \dodoi{10.1126/sciadv.1500686}

\bibitem[{Schwarz(1978)}]{schwarz1978}
Schwarz, G. 1978, Ann. Statist., 6, 461, \dodoi{10.1214/aos/1176344136}

\bibitem[{{Scott} \& {Leahy}(1999)}]{scott1999}
{Scott}, D.~M., \& {Leahy}, D.~A. 1999, \apj, 510, 974, \dodoi{10.1086/306631}

\bibitem[{{Singh} {et~al.}(2014){Singh}, {Tandon}, {Agrawal}, {Antia},
  {Manchanda}, {Yadav}, {Seetha}, {Ramadevi}, {Rao}, {Bhattacharya}, {Paul},
  {Sreekumar}, {Bhattacharyya}, {Stewart}, {Hutchings}, {Annapurni}, {Ghosh},
  {Murthy}, {Pati}, {Rao}, {Stalin}, {Girish}, {Sankarasubramanian},
  {Vadawale}, {Bhalerao}, {Dewangan}, {Dedhia}, {Hingar}, {Katoch}, {Kothare},
  {Mirza}, {Mukerjee}, {Shah}, {Shah}, {Mohan}, {Sangal}, {Nagabhusana},
  {Sriram}, {Malkar}, {Sreekumar}, {Abbey}, {Hansford}, {Beardmore}, {Sharma},
  {Murthy}, {Kulkarni}, {Meena}, {Babu}, \& {Postma}}]{singh2014}
{Singh}, K.~P., {Tandon}, S.~N., {Agrawal}, P.~C., {et~al.} 2014, in \procspie,
  Vol. 9144, Space Telescopes and Instrumentation 2014: Ultraviolet to Gamma
  Ray, 91441S

\bibitem[{{Smith} {et~al.}(2018){Smith}, {Mushotzky}, {Boyd}, {Malkan},
  {Howell}, \& {Gelino}}]{smith2018}
{Smith}, K.~L., {Mushotzky}, R.~F., {Boyd}, P.~T., {et~al.} 2018, \apj, 857,
  141, \dodoi{10.3847/1538-4357/aab88d}

\bibitem[{Staubert {et~al.}(2017)Staubert, Klochkov, F\"urst, Wilms,
  Rothschild, \& Harrison}]{staubert_2017}
Staubert, R., Klochkov, D., F\"urst, F., {et~al.} 2017, {Astronomy \&
  Astrophysics}, 606, L13, \dodoi{10.1051/0004-6361/201731927}

\bibitem[{{Stevens} \& {Uttley}(2016)}]{stevensuttley2016}
{Stevens}, A.~L., \& {Uttley}, P. 2016, \mnras, 460, 2796,
  \dodoi{10.1093/mnras/stw1093}

\bibitem[{{Strohmayer} \& {Watts}(2005)}]{strohmayer2005}
{Strohmayer}, T.~E., \& {Watts}, A.~L. 2005, \apjl, 632, L111,
  \dodoi{10.1086/497911}

\bibitem[{{Tananbaum} {et~al.}(1972){Tananbaum}, {Gursky}, {Kellogg},
  {Levinson}, {Schreier}, \& {Giacconi}}]{tananbaum1972}
{Tananbaum}, H., {Gursky}, H., {Kellogg}, E.~M., {et~al.} 1972, \apjl, 174,
  L143, \dodoi{10.1086/180968}

\bibitem[{Taylor(1992)}]{Taylor92}
Taylor, J.~H. 1992, Philosophical Transactions: Physical Sciences and
  Engineering, 341, 117

\bibitem[{{Tetarenko} {et~al.}(2019){Tetarenko}, {Casella}, {Miller-Jones},
  {Sivakoff}, {Tetarenko}, {Maccarone}, {Gandhi}, \&
  {Eikenberry}}]{tetarenko2019}
{Tetarenko}, A.~J., {Casella}, P., {Miller-Jones}, J.~C.~A., {et~al.} 2019,
  \mnras, 484, 2987, \dodoi{10.1093/mnras/stz165}

\bibitem[{{The Astropy Collaboration} {et~al.}(2018){The Astropy
  Collaboration}, {Price-Whelan}, {Sip{\H o}cz}, {G{\"u}nther}, {Lim},
  {Crawford}, {Conseil}, {Shupe}, {Craig}, {Dencheva}, {Ginsburg},
  {VanderPlas}, {Bradley}, {P{\'e}rez-Su{\'a}rez}, {de Val-Borro}, {Aldcroft},
  {Cruz}, {Robitaille}, {Tollerud}, {Ardelean}, {Babej}, {Bachetti}, {Bakanov},
  {Bamford}, {Barentsen}, {Barmby}, {Baumbach}, {Berry}, {Biscani}, {Boquien},
  {Bostroem}, {Bouma}, {Brammer}, {Bray}, {Breytenbach}, {Buddelmeijer},
  {Burke}, {Calderone}, {Cano Rodr{\'{\i}}guez}, {Cara}, {Cardoso},
  {Cheedella}, {Copin}, {Crichton}, {D{\'A}vella}, {Deil}, {Depagne},
  {Dietrich}, {Donath}, {Droettboom}, {Earl}, {Erben}, {Fabbro}, {Ferreira},
  {Finethy}, {Fox}, {Garrison}, {Gibbons}, {Goldstein}, {Gommers}, {Greco},
  {Greenfield}, {Groener}, {Grollier}, {Hagen}, {Hirst}, {Homeier}, {Horton},
  {Hosseinzadeh}, {Hu}, {Hunkeler}, {Ivezi{\'c}}, {Jain}, {Jenness}, {Kanarek},
  {Kendrew}, {Kern}, {Kerzendorf}, {Khvalko}, {King}, {Kirkby}, {Kulkarni},
  {Kumar}, {Lee}, {Lenz}, {Littlefair}, {Ma}, {Macleod}, {Mastropietro},
  {McCully}, {Montagnac}, {Morris}, {Mueller}, {Mumford}, {Muna}, {Murphy},
  {Nelson}, {Nguyen}, {Ninan}, {N{\"o}the}, {Ogaz}, {Oh}, {Parejko}, {Parley},
  {Pascual}, {Patil}, {Patil}, {Plunkett}, {Prochaska}, {Rastogi}, {Reddy
  Janga}, {Sabater}, {Sakurikar}, {Seifert}, {Sherbert}, {Sherwood-Taylor},
  {Shih}, {Sick}, {Silbiger}, {Singanamalla}, {Singer}, {Sladen}, {Sooley},
  {Sornarajah}, {Streicher}, {Teuben}, {Thomas}, {Tremblay}, {Turner},
  {Terr{\'o}n}, {van Kerkwijk}, {de la Vega}, {Watkins}, {Weaver}, {Whitmore},
  {Woillez}, \& {Zabalza}}]{astropy}
{The Astropy Collaboration}, {Price-Whelan}, A.~M., {Sip{\H o}cz}, B.~M.,
  {et~al.} 2018, ArXiv e-prints.
\newblock \doarXiv{1801.02634}

\bibitem[{{Timmer} \& {Koenig}(1995)}]{timmer1995}
{Timmer}, J., \& {Koenig}, M. 1995, \aap, 300, 707

\bibitem[{{Uttley} {et~al.}(2014){Uttley}, {Cackett}, {Fabian}, {Kara}, \&
  {Wilkins}}]{uttley2014}
{Uttley}, P., {Cackett}, E.~M., {Fabian}, A.~C., {Kara}, E., \& {Wilkins},
  D.~R. 2014, \aapr, 22, 72, \dodoi{10.1007/s00159-014-0072-0}

\bibitem[{{Uttley} \& {McHardy}(2001)}]{uttley2001}
{Uttley}, P., \& {McHardy}, I.~M. 2001, \mnras, 323, L26,
  \dodoi{10.1046/j.1365-8711.2001.04496.x}

\bibitem[{{van der Klis}(1989)}]{vanderklis1989}
{van der Klis}, M. 1989, in NATO Advanced Science Institutes (ASI) Series C,
  Vol. 262, NATO Advanced Science Institutes (ASI) Series C, ed.
  H.~{{\"O}gelman} \& E.~P.~J. {van den Heuvel}, 27

\bibitem[{van~der Walt {et~al.}(2011)van~der Walt, Colbert, \&
  Varoquaux}]{numpy}
van~der Walt, S., Colbert, S.~C., \& Varoquaux, G. 2011, Computing in Science
  Engineering, 13, 22, \dodoi{10.1109/MCSE.2011.37}

\bibitem[{{Vaughan} {et~al.}(1994){Vaughan}, {van der Klis}, {Lewin}, {Wijers},
  {van Paradijs}, {Dotani}, \& {Mitsuda}}]{Vaughanetal94}
{Vaughan}, B., {van der Klis}, M., {Lewin}, W.~H.~G., {et~al.} 1994, \apj, 421,
  738, \dodoi{10.1086/173686}

\bibitem[{{Vaughan} \& {Nowak}(1997)}]{vaughan1997}
{Vaughan}, B.~A., \& {Nowak}, M.~A. 1997, \apjl, 474, L43,
  \dodoi{10.1086/310430}

\bibitem[{{Vaughan}(2010)}]{vaughan2010}
{Vaughan}, S. 2010, \mnras, 402, 307, \dodoi{10.1111/j.1365-2966.2009.15868.x}

\bibitem[{{Vaughan} {et~al.}(2003){Vaughan}, {Edelson}, {Warwick}, \&
  {Uttley}}]{Vaughanetal03}
{Vaughan}, S., {Edelson}, R., {Warwick}, R.~S., \& {Uttley}, P. 2003, \mnras,
  345, 1271, \dodoi{10.1046/j.1365-2966.2003.07042.x}

\bibitem[{{Walton} {et~al.}(2018){Walton}, {Bachetti}, {F{\"u}rst}, {Barret},
  {Brightman}, {Fabian}, {Grefenstette}, {Harrison}, {Heida}, {Kennea},
  {Kosec}, {Lau}, {Madsen}, {Middleton}, {Pinto}, {Steiner}, \&
  {Webb}}]{walton2018}
{Walton}, D.~J., {Bachetti}, M., {F{\"u}rst}, F., {et~al.} 2018, \apj, 857, L3,
  \dodoi{10.3847/2041-8213/aabadc}

\bibitem[{{Wang-Ji} {et~al.}(2018){Wang-Ji}, {Garc{\'\i}a}, {Steiner},
  {Tomsick}, {Harrison}, {Bambi}, {Petrucci}, {Ferreira}, {Chakravorty}, \&
  {Clavel}}]{WangJietal18}
{Wang-Ji}, J., {Garc{\'\i}a}, J.~A., {Steiner}, J.~F., {et~al.} 2018, \apj,
  855, 61, \dodoi{10.3847/1538-4357/aaa974}

\bibitem[{{Watts} \& {Strohmayer}(2006)}]{watts2006}
{Watts}, A.~L., \& {Strohmayer}, T.~E. 2006, \apjl, 637, L117,
  \dodoi{10.1086/500735}

\bibitem[{{Wilkinson} \& {Uttley}(2009)}]{WilkinsonUttley09}
{Wilkinson}, T., \& {Uttley}, P. 2009, \mnras, 397, 666,
  \dodoi{10.1111/j.1365-2966.2009.15008.x}

\bibitem[{{Yamaoka} {et~al.}(2010){Yamaoka}, {Sugizaki}, {Nakahira}, {Mihara},
  {Kohama}, {Nakagawa}, {Yamamoto}, {Matsuoka}, {Kawasaki}, {Ueno}, {Tomida},
  {Suzuki}, {Ishikawa}, {Kawai}, {Morii}, {Sugimori}, {Yoshida}, {Tsunemi},
  {Kimura}, {Negoro}, {Nakajima}, {Ishiwata}, {Miyoshi}, {Ozawa}, {Ueda},
  {Isobe}, {Eguchi}, {Hiroi}, \& {Daikyuji}}]{Yamaokaetal10}
{Yamaoka}, K., {Sugizaki}, M., {Nakahira}, S., {et~al.} 2010, The Astronomer's
  Telegram, 2380

\bibitem[{{Zdziarski} {et~al.}(1998){Zdziarski}, {Poutanen}, {Mikolajewska},
  {Gierlinski}, {Ebisawa}, \& {Johnson}}]{Zdziarskietal98}
{Zdziarski}, A.~A., {Poutanen}, J., {Mikolajewska}, J., {et~al.} 1998, \mnras,
  301, 435, \dodoi{10.1046/j.1365-8711.1998.02021.x}

\bibitem[{{Zhang} {et~al.}(2016){Zhang}, {Feroci}, {Santangelo}, {Dong},
  {Feng}, {Lu}, {Nandra}, {Wang}, {Zhang}, {Bozzo}, {Brandt}, {De Rosa}, {Gou},
  {Hernanz}, {van der Klis}, {Li}, {Liu}, {Orleanski}, {Pareschi}, {Pohl},
  {Poutanen}, {Qu}, {Schanne}, {Stella}, {Uttley}, {Watts}, {Xu}, {Yu}, {in 't
  Zand}, {Zane}, {Alvarez}, {Amati}, {Baldini}, {Bambi}, {Basso},
  {Bhattacharyya S.}, {}, {Belloni}, {Bellutti}, {Bianchi}, {Brez}, {Bursa},
  {Burwitz}, {Budtz-J{\o}rgensen}, {Caiazzo}, {Campana}, {Cao}, {Casella},
  {Chen}, {Chen}, {Chen}, {Chen}, {Chen}, {Chen}, {Civitani}, {Coti Zelati},
  {Cui}, {Cui}, {Dai}, {Del Monte}, {de Martino}, {Di Cosimo}, {Diebold},
  {Dovciak}, {Donnarumma}, {Doroshenko}, {Esposito}, {Evangelista}, {Favre},
  {Friedrich}, {Fuschino}, {Galvez}, {Gao}, {Ge}, {Gevin}, {Goetz}, {Han},
  {Heyl}, {Horak}, {Hu}, {Huang}, {Huang}, {Hudec}, {Huppenkothen}, {Israel},
  {Ingram}, {Karas}, {Karelin}, {Jenke}, {Ji}, {Korpela}, {Kunneriath},
  {Labanti}, {Li}, {Li}, {Li}, {Liang}, {Limousin}, {Lin}, {Ling}, {Liu},
  {Liu}, {Liu}, {Lu}, {Lund}, {Lai}, {Luo}, {Luo}, {Ma}, {Mahmoodifar},
  {Marisaldi}, {Martindale}, {Meidinger}, {Men}, {Michalska}, {Mignani},
  {Minuti}, {Motta}, {Muleri}, {Neilsen}, {Orlandini}, {Pan}, {Patruno},
  {Perinati}, {Picciotto}, {Piemonte}, {Pinchera}, {Rachevski A.}, {Rapisarda},
  {Rea}, {Rossi}, {Rubini}, {Sala}, {Shu}, {Sgro}, {Shen}, {Soffitta}, {Song},
  {Spandre}, {Stratta}, {Strohmayer}, {Sun}, {Svoboda}, {Tagliaferri},
  {Tenzer}, {Hong}, {Taverna}, {Torok}, {Turolla}, {Vacchi}, {Wang}, {Walton},
  {Wang}, {Wang}, {Wang}, {Wang}, {Weng}, {Wilms}, {Winter}, {Wu}, {Wu},
  {Xiong}, {Xu}, {Xue}, {Yan}, {Yang}, {Yang}, {Yang}, {Yuan}, {Yuan}, {Yuan},
  {Zampa}, {Zampa}, {Zdziarski}, {Zhang}, {Zhang}, {Zhang}, {Zhang}, {Zhang},
  {Zhang}, {Zheng}, {Zhou}, \& {Zhou X.~L.}}]{extp16}
{Zhang}, S.~N., {Feroci}, M., {Santangelo}, A., {et~al.} 2016, in \procspie,
  Vol. 9905, Space Telescopes and Instrumentation 2016: Ultraviolet to Gamma
  Ray, 99051Q

\end{thebibliography}
\bibliographystyle{aasjournal}


\end{document}